\documentclass[11pt,a4paper]{article}
\usepackage[T1]{fontenc}
\usepackage[round]{natbib}
\usepackage{newtxtext,newtxmath}
\usepackage[cm]{fullpage}
\usepackage[pdftex]{graphicx}
\usepackage{url}
\usepackage{xcolor}
\usepackage[hidelinks,breaklinks]{hyperref}

\newcommand{\firstuse}[1]{\emph{#1}} 
\newcommand{\ie}{i.e.\ }

\newcommand{\ignore}[1]{}
\newcommand{\unix}[1]{\texttt{#1}} 
\newcommand{\email}[1]{\texttt{#1}}
\providecommand{\keywords}[1]{{\small\textbf{\textit{Keywords---}} #1}}

\makeatletter
\let\@fnsymbol\@arabic
\makeatother

\begin{document}

\title{Geodesic cycle length distributions in fictional character networks}

\author{Alex Stivala\thanks{Universit\`a della  Svizzera italiana, Via Giuseppe Buffi 13, 6900 Lugano, Switzerland. Email: \email{alexander.stivala@usi.ch}}}

\maketitle

\begin{abstract}
  A geodesic cycle in a graph is a cycle with no shortcuts, so that
  the shortest path between any two nodes in the cycle is the path
  along the cycle itself. A recently published paper used random graph
  models to investigate the geodesic cycle length distributions of a
  unique set of delusional social networks, first examined in an
  earlier work, as well as some other publicly available social
  networks. Here I test the hypothesis, suggested in the former work,
  that fictional character networks, and in particular those
  from works by a single author, might have geodesic cycle length
  distributions which are extremely unlikely under random graph
  models, as the delusional social networks do. The results do not
  show any support for this hypothesis. In addition, the recently
  published work is reproduced using a method for counting geodesic
  cycles exactly, rather than the approximate method used originally.
  The substantive conclusions of that work are unchanged, but some
  differences in the results for particular networks are described.
\end{abstract}


\section{Introduction}


In order to investigate the hypothesis that people tend to
conceive of social networks in local and spatial terms,
\citet{martin17} used random graphs to analyze the structure of a
unique set of social networks. These three networks are social
networks of alternative personalities described by a patient,
``Patricia'', undergoing therapy for dissociative identity disorder,
whose drawings of these networks are reproduced in \citet{david96}.
One of the most striking findings in \cite{martin17} is the presence
in one of these networks of a large ``hollow ring'', a cycle with no
``shortcuts'', which is much larger than those expected under the
random graph models considered.  As discussed in \citet{stivala20a}
and recapitulated below, this ``hollow ring'' is in fact a
\firstuse{geodesic cycle} in graph theory terms.

In \citet{martin17}, the random graph null model used was the
$dk$-series \citep{mahadevan06,orsini15} family of graph
distributions, and only the distribution of the largest geodesic cycle
lengths is considered. \citet{stivala20a} extends this work by using
exponential random graph models (ERGM) \citep{lusher13} as well as the
$dk$-series random graphs, and also considering the geodesic cycle
length distributions, in addition to the maximum geodesic cycle
lengths. That work also examined a selection of other publicly
available social networks, as well as the original ``Patricia''
networks, finding that, of the networks examined, only the
``Patricia'' networks seem to have geodesic cycle lengths that are not
reproduced by random graph models.

Commentary on \citet{stivala20a} mostly concerned the
[mis-]interpretation of ERGM parameters and the trend (or otherwise) in
the social networks field to be moving towards a monoculture of using
parametric models, and in particular ERGMs, for everything
\citep{martin20,stivala20b}. What has not yet been addressed, is to
test the hypothesis suggested in \citet{stivala20a}, that fictional
character networks (see \citet{labatut19} for an overview), and
specifically those with only a single author (unlike the one fictional
character network examined in that work), might have anomalous
geodesic cycle length distributions, like the ``Patricia'' networks.

The primary purpose of this work is to test this hypothesis, by
examining how well the geodesic cycle length distributions in some
fictional character networks --- mostly from works by a single author
--- can be reproduced by random graph models, specifically the
$dk$-series random graphs, and, in some cases, exponential random graph
models.

Further, it was implicitly assumed in \citet{stivala20a} that the
method used there for counting geodesic cycles was exact. In fact,
however, that method is only an approximation (more precisely, a lower
bound). And so a second purpose of this work is to reproduce all
of the results in \citet{stivala20a} and compare with a method that
counts geodesic cycles exactly.

\section{Geodesic cycles}
\label{sec:filtercyclesgeodesic}

In this and the following sections, only undirected graphs are
considered.

As discussed in \citet{stivala20a}, the structure that
\citet{martin17} describes as a ``hollow ring'' --- a cycle with no
``shortcuts'', so that the shortest path between any two nodes in the
cycle is along the cycle itself --- is known in graph theory as a
\firstuse{geodesic cycle} \citep{negami86,li18b}. Another way of
putting this is that a geodesic cycle is a cycle such that, for any
two vertices on the cycle, the distance between them in the cycle is
equal to their distance in the graph \citep{amaldi09,lokshtanov09}.

As \citet[p.~279]{mitchell74} writes of the ``proliferation of
concepts and terms'' in social network analysis, it is a
''terminological jungle, in which any newcomer may plant a tree''
(\citet[p.~8]{barnes72} as quoted in \citet[p.~279]{mitchell74}), and
as noted in \citet{stivala20a}, a geodesic cycle has also been termed
an \firstuse{isometric cycle} \citep{lokshtanov09}, and an
\firstuse{atomic cycle} \citep{gashler12}.  Adding yet another tree to
the terminological jungle, an isometric cycle has also been termed a
\firstuse{short cycle} \citep{klemm06}.

Unlike geodesic cycle, \firstuse{chordless cycle} is a well-known
standard term in graph theory, although there are conflicting
conventions as to whether or not cycles of length three are included
in the definition \citep{mathworld}. A \firstuse{chord} is an edge
connecting two otherwise non-adjacent nodes in a cycle, and a
chordless cycle is a cycle that has no chords. A chordless cycle is
also known as an \firstuse{induced cycle} or sometimes a
\firstuse{hole}, for chordless cycles of length at least four.
Every geodesic cycle is chordless, but not every chordless cycle
is geodesic. An illustrated example of these definitions is given
in \citet[Fig.~1]{stivala20a}.

In \citet{stivala20a}, the \texttt{find\_large\_atomic\_cycle}
algorithm \citep{gashler11,gashler12} was used to count geodesic
cycles. However, despite the implicit assumption in that work that
this algorithm enumerates geodesic cycles (that is, finds all of
them), in fact, this is not guaranteed. The proof of correctness in
\citet{gashler12} proves that any cycle it returns is geodesic, and
that it will find at least one of the specified minimum length
if it exists.\footnote{However, as can be seen, from, for example,
Fig.~\ref{fig:dolphins_geodesiccycle_lengthdist}, where
\texttt{find\_large\_atomic\_cycle} does not find a geodesic cycle of
length 7, four of which are found by the method described in this
section, either this proof is not correct or there is an error in the
implementation.}  However, despite being described as ``iterating over
all the atomic cycles in the graph'' and that ``atomic cycles in the
graph are enumerated'' \citep[p.~6]{gashler12}, the proof does not
show that all atomic (geodesic) cycles are enumerated (or iterated
over), and in fact it is not guaranteed to do so. Hence it does not
exactly count geodesic cycles, but only approximates the count. More
precisely, it gives a lower bound on the number of geodesic cycles
of each length.
This is perfectly suitable for the purpose for which this algorithm
was designed (finding such cycles and ``cutting'' them to improve the
performance of a machine learning algorithm), but if we are going to
count geodesic cycles, we would like an algorithm that can be
guaranteed to do so exactly (even if only to get an idea of how close
the, more easily obtained, lower bound is to the exactly value).

In the
case of the three ``Patricia'' networks considered in
\citet{martin17,stivala20a}, in fact the counts are exact, which was
verified by manual inspection.  However this is not the case for all
networks, as will be shown below (see
Section~\ref{sec:results_reproduction} and \ref{sec:reproduction}).

In this work, in order to count geodesic cycles exactly, an algorithm
\unix{isCycleGeodesic} is used. This algorithm tests if a given cycle
in a graph is geodesic directly by the definition. That is, for every
pair of nodes in the cycle, it tests if the distance between those
nodes in the cycle is equal to the shortest path distance (geodesic)
between them in the graph. It is a precondition of this function that
the cycle is specified as a list of nodes in order along the cycle, so
the cycle distance is efficiently computed as $\min\{\lvert i-j\rvert,
k - \lvert i-j \rvert \}$, where $i$ and $j$ are the indices of the
nodes in the cycle, $i,j \in \{ 0 \ldots k-1 \}$, and $k$ is the length
of the cycle. The distance matrix (giving shortest path length for
every pair of nodes) for the graph can be pre-computed, or, more
efficiently, a single-source shortest paths algorithm can be used on
demand, and the results memoized for reuse.  This algorithm is
implemented by the \unix{filterCyclesGeodesic.py} Python script, which
takes a graph and, on the input stream, a list of cycles in the graph,
and writes to the output stream only those cycles from the input which are
geodesic.

Note that, because all geodesic cycles are chordless, it is more
efficient to give as input to this algorithm only chordless cycles,
which can be enumerated more efficiently than having to enumerate all
cycles \citep{uno14}.

\section{Exponential random graph models and \texorpdfstring{$dk$}{dk}-series random graphs}

As in \citet{stivala20a}, I use two different families of random graphs:
exponential random graph models (ERGMs), and the $dk$-series family of random
graphs.

ERGMs are a random graph model, widely used in the social
sciences to model social networks
\citep{lusher13,amati18,koskinen20}. An ERGM is a
probability distribution over graphs, where the probability of a graph
is a function of a set of graph statistics with their associated
parameters.  The statistics are counts of ``configurations'', which
represent local structures in the graph, such as edges, stars, or
triangles. These statistics need not be purely structural, but can
also include nodal attributes, in order to model homophily, for
example. Given an observed graph, the parameters corresponding to the
configurations in the model can be estimated by maximum likelihood,
with their sign and significance indicating over- or
under- representation, given all the other parameters in the model, of
the corresponding configuration.

In this work, I do not make any interpretation of the estimated
parameters, but simply use them to simulate a set of graphs from the
estimated model, for use as a null model in comparing the observed
geodesic cycle distributions to those generated from the model.

In contrast, the $dk$-series random graphs \citep{orsini15} are a
series of random graph null models of increasing complexity, forming a
nested hierarchy. For each level of the hierarchy, an ensemble of
random graphs are generated, for which particular statistics are fixed
at the values in the observed graph.  The simplest level of the
hierarchy is $0k$, in which the density is fixed. This is just
the Erd{\H{o}}s--R{\'e}nyi random graph model.  This is followed by
$1k$, in which the degree distribution is also fixed, again, a
well-known random graph null model \citep{newman01}, used for
``motif'' detection in biological networks, for example
\citep{milo02}. The next level is $2k$, which also fixes the joint
degree distribution, thus reproducing the degree correlations
(assortativity) of the observed graph.  For practical use in
generating ensembles of graphs that resemble empirical networks in
relevant statistics, \citet{orsini15} define the next two
distributions in the hierarchy as $2.1k$ and $2.5k$. These are the
$2k$-random graphs with fixed values of average local clustering and
average local clustering by degree (clustering spectrum),
respectively.

An advantage of the $dk$-series approach is that we do not need to
contend with the potential difficulties in estimating ERGM parameters,
where for some networks, such as most of the fictional character
networks considered here, it can be difficult or perhaps even impossible
(without adding a lot of extra constraints, such as fixing the
presence or absence of particular ties) to find a converged
model. Further discussion of the relative merits and shortcomings of
these approaches can be found in \citet{martin17,stivala20a,martin20,stivala20b}.

\section{Data}

The nine networks used in \citet{stivala20a} include the three
``Patricia'' networks from \citet{david96} originally used in
\citet{martin17}, as well as six other networks from a variety of
domains
\citep{badhessian,weissman19,leavitt14,lusseau03,lazega99,lazega01,zachary77,kapferer72,mastrandrea15},
as described in \citet{stivala20a}.

The seven fictional character networks used in this work consist of
four chapter co-occurrence networks from classic novels, one character
interaction network from the \textit{Star Wars} movie series, and two
other types of character interaction networks from the \textit{Harry
  Potter} books. The four chapter co-occurrence networks considered are
from \textit{Anna Karenina} by Leo Tolstoy, \textit{David Copperfield}
by Charles Dickens, \textit{Adventures of Huckleberry Finn} by Mark
Twain, and \textit{Les Mis\'{e}rables} by Victor Hugo.

The Anna Karenina, David Copperfield, and Huckleberry Finn character
interaction networks \citep{knuth93} were downloaded from
\url{https://people.sc.fsu.edu/~jburkardt/datasets/sgb/sgb.html}
[accessed 1 September 2019]. Data in the Stanford GraphBase format
 \citep{knuth93} was parsed using Python code adapted from the
\unix{charnet} project \citep{holanda19} from
\url{https://github.com/ajholanda/charnet}.
These networks are illustrated in Figures
\ref{fig:annakarenina_network}, \ref{fig:davidcopperfield_network}, and
\ref{fig:huckleberryfinn_network}.
The Les Mis\'{e}rables character interaction network \citep{knuth93}
was downloaded from
\url{http://www-personal.umich.edu/~mejn/netdata/lesmis.zip} [accessed
  10 July 2019]. This network is illustrated in
Fig.~\ref{fig:lesmiserables_network}.

The two \textit{Harry Potter} networks are ``Dumbledore's Army'',
and the peer support network for the first six Harry Potter books.
The ``Dumbledore's Army'' network \citep{everton22} from the Harry
Potter books was downloaded from
\url{https://core-dna.netlify.app/publication/harry_potter_dumbledores_army}
[accessed 13 January 2023]. A binary network was constructed by
including only edges (representing trust or mutual understanding)
with a tie strength of 4 or 5, resulting in the network shown in
\citet[Fig.~2]{everton22} and Fig.~\ref{fig:dumbledoresarmy_network}.

The Harry Potter peer support network \citep{bossaert13} for the
first six Harry Potter books was downloaded from
\url{http://www.stats.ox.ac.uk/~snijders/siena/HarryPotterData.html}
[accessed 12 July 2019]. The network is directed, with a
directed edge $(i,j)$ indicating that character $i$ gives peer
support to character $j$. The network was converted to undirected
by changing any directed edge to an undirected edge (mutual
directed edges result in a single undirected edge). This network
also contains node attributes for school year, gender, and
wizarding house, and is illustrated in Fig.~\ref{fig:harrypotter_network}.

The ``merged'' Star Wars character interaction network was used,
meaning Anakin and Darth Vader are the same node
\citep{gabasova15,gabasova16}.  This network is illustrated in
Fig.~\ref{fig:starwars_network}.

Summary statistics of these networks are shown in
Table~\ref{tab:graph_stats_summary}.

\begin{table}
  \caption{Summary statistics of the networks.}
  \label{tab:graph_stats_summary}
  \begin{center}
\begin{tabular}{lrrrrrrr}
\hline
Network & N  &   Components &  Mean   &    Density & Clustering  & Assortativity &  Mean \\
        &    &              &  degree &            & coefficient & coefficient   & path length \\
\hline
Anna Karenina & 138 & 1 & 7.14 & 0.05215 & 0.26827 & -0.35010 & 2.45\\
David Copperfield & 87 & 1 & 9.33 & 0.10853 & 0.35124 & -0.25814 & 1.95\\
Huckleberry Finn & 74 & 3 & 8.14 & 0.11144 & 0.48756 & -0.17297 & 2.14\\
Les Mis\'{e}rables & 77 & 1 & 6.60 & 0.08681 & 0.49893 & -0.16523 & 2.64\\
Dumbledore's Army & 29 & 1 & 4.90 & 0.17488 & 0.54404 & 0.19288 & 2.57\\
Harry Potter & 64 & 29 & 3.62 & 0.05754 & 0.52844 & -0.19461 & 1.90\\
Star Wars & 111 & 2 & 8.00 & 0.07273 & 0.35099 & -0.20863 & 2.61\\
\hline
\end{tabular}
  \end{center}
  \parbox{\textwidth}{\footnotesize All networks are undirected.  ``Clustering
    coefficient'' is the global clustering coefficient (transitivity),
    and ``Assortativity coefficient'' is the degree
    assortativity. All statistics were computed using the \unix{igraph}
    \citep{csardi06} library in R \citep{R-manual}.}
\end{table}

\section{Methods}

ERGM models were estimated, and networks simulated from the models,
with the \unix{ergm} R package in the \unix{statnet} software suite
\citep{handcock08,hunter08,hummel12,statnet,ergm,ergm4}.  Networks
were simulated from the $dk$-series random graph models using the
\unix{RandNetGen}
software\footnote{\url{https://github.com/polcolomer/RandNetGen}}
\citep{mahadevan06,colomer13,colomer14,orsini15}.  The
\unix{RandNetGen} program does not simulate graphs from the $0k$
distribution (Erd{\H{o}}s--R{\'e}nyi random graphs), so, as in
\citet{martin17,stivala20a}, the \unix{igraph} R package
\citep{csardi06} was used to simulate these networks.

For each ERGM model or $dk$-series distribution, 100 networks are
simulated, and the box plots, generated with the \unix{ggplot2} R
package \citep{ggplot2} represent data from these 100 networks.

For the networks considered in \citet{stivala20a}, the same sets of
simulated networks from that work were used. Further
details of the ERGM models in that work can be found in
\citet{stivala20a,stivala20b}.

To count geodesic cycles in a network, the \unix{CYPATH}
software\footnote{\url{http://research.nii.ac.jp/~uno/code/cypath.html}}
\citep{uno14} was used to generate all of the chordless cycles, which
were then given as input to the \unix{filterCyclesGeodesic.py} script (see
Section~\ref{sec:filtercyclesgeodesic}), implemented in Python with
the \unix{igraph} library, which outputs only those cycles that are geodesic.

To find the longest geodesic cycle in a graph, Algorithm 4.1, ``LIC --
Longest Isometric Cycle'' from \citet{lokshtanov09} was implemented in
Python in the \unix{longestIsometricCycle.py} script. It is important
to note that, as described in \citet{catrina21}, Lemma~3.6 of
\citet{lokshtanov09} is erroneous, and that therefore the algorithm is
only correct for even-length cycles. For odd-length cycles, the
conditions of this Lemma may be met for a particular cycle of length
$k$ (for an odd value of $k$), and yet there is no cycle of length $k$
in the graph, and therefore the algorithm will incorrectly find that
there is a geodesic cycle of length $k$. There is an example showing
this case in \citet[Fig.~5]{catrina21}, and it also occurs, for
example, in the ``Patricia'' graph for 1990, in which the conditions
of \citet[Lemma~3.6]{lokshtanov09} are met for a geodesic cycle of
length 11, and yet there is no such cycle: the longest geodesic cycle
is of length 10 (see \citet{martin17,stivala20a}, and
Fig.~\ref{fig:patricia1990_geodesiccycle_lengthdist}).  In order to
have the algorithm work correctly for both odd and even values of $k$,
the auxiliary bipartite graph construction described by
\citet[Observation~5.4]{catrina21} was implemented in
\unix{longestIsometricCycle.py}. It is an assumption of this algorithm
that the input graph is connected, so the implementation handles
graphs that are not connected by running the algorithm on each
connected component, and returning the size of the largest geodesic
cycle found in any component. This is the longest geodesic cycle in
the graph, since any cycle must be entirely contained within a single
connected component.

For each set of 100 simulated networks, the script to count geodesic
cycles was run with an elapsed time limit of 48 hours, on an Intel
Xeon E5-2650 v3 2.30GHz processor on a Linux compute cluster node.
The iteration over the 100 simulated networks was done serially, that
is, one network at a time. (This serial counting process for each set
of 100 networks was done in parallel over the networks simulated from
different distributions, however). If the time limit was reached, then
only the results for the networks for which the counting could be
completed within the time limit are included. In such cases, where
this number was less than 100, this is noted in the relevant figure
caption. In some cases, for the $0k$-distribution, the count cannot be
completed within the time limit for even the first network in the set
of simulated networks, and so there are no results, and the results
for this distribution are omitted.

The lengths of the largest geodesic cycles, as shown in the plots in
Section~\ref{sec:fiction_results} and \ref{sec:reproduction}, are
simply the largest geodesic cycles in the distributions, computed as
described above, that have a nonzero count. In order to further verify
that these are correct, the \unix{longestIsometricCycle.py} script
described above was used to find the length of the longest geodesic
(\ie isometric) cycle in each network, and I verified that these were
the same as those found from \unix{filterCyclesGeodesic.py}

\section{Results and discussion}

\subsection{Reproducing earlier approximate results with exact counts}
\label{sec:results_reproduction}

In \ref{sec:reproduction}, the results from \citet{stivala20a}, where
geodesic cycles were counted only approximately (lower bound) using
the \texttt{find\_large\_atomic\_cycle} algorithm
\citep{gashler11,gashler12} are plotted on the same plots as those
where the geodesic cycles are counted exactly, as described
Section~\ref{sec:filtercyclesgeodesic}.  For Patricia's 1990
(Fig.~\ref{fig:patricia1990_geodesiccycle_lengthdist}) and 1992
(Fig.~\ref{fig:patricia1992_geodesiccycle_lengthdist}) networks, the
results are not any different from those described in
\citet{stivala20a}: in the 1990 network, the largest geodesic cycle
(length 10) is extremely unlikely under any of the models (ERGM or
$dk$-series), while in the 1992 network, that the largest geodesic
cycle is only of length 4 is extremely unlikely under any of the
models (in this case, larger maximum geodesic cycle lengths are
expected).  The results for the geodesic cycle length distributions
are also unchanged, with the fit being poor for the 1990 and 1992
networks, but acceptable, in the case of the ERGM only, for the 1993
network (Fig.~\ref{fig:patricia1993_geodesiccycle_lengthdist}).  The
situation for maximum geodesic cycle lengths in the 1993 network is
slightly different, with the exact counting showing that the largest
geodesic cycle (length 8) is more clearly unlikely than is apparent
when the approximate (lower bound) method was used.  Along with
Fig.~\ref{fig:patricia1993_geodesiccycle_more_lengthdist}, this shows
that, just as discussed in \citet{stivala20a}, the ERGM shows a better
fit to the geodesic cycle length distribution in this network than any
of the $dk$-series distributions.

For all three ``Patricia'' networks, the observed maximum geodesic
cycle length is found to be the same by both methods; indeed the fact
that the \texttt{find\_large\_atomic\_cycle} method found the correct largest
geodesic cycle lengths was verified manually on these networks.

Figures \ref{fig:greysanatomy_geodesiccycle_lengthdist} --
\ref{fig:highschoolfriendship_geodesiccycle_lengthdist} show the
results for the other networks considered in \citet{stivala20a}. When
using the approximate (lower bound) counting method, the ERGM shows a
good fit for the largest geodesic cycle size; however when using the
exact counting method, this is not the case for the dolphin social
network (Fig.~\ref{fig:dolphins_geodesiccycle_lengthdist}) or the
Kapferer tailor shop network
(Fig.~\ref{fig:kapferertailorshop_geodesiccycle_lengthdist}).  In
\citet[p.~53]{stivala20a} it was stated that the ERGM shows an
acceptable fit to the geodesic cycle length distributions for all the
non-''Patricia'' networks considered, as does the $dk$-series $2.5k$
distribution (the latter with the exception of the Grey's Anatomy and
high school friendship networks). However when using the exact
counting method, we can see that this is not true for the
dolphin social network
(Fig.~\ref{fig:dolphins_geodesiccycle_lengthdist}), the Lazega law firm
network (Fig.~\ref{fig:lawfirm_geodesiccycle_lengthdist}), the Zachary
karate club network
(Fig.~\ref{fig:zacharykarateclub_geodesiccycle_lengthdist}), and the
Kapferer tailor shop network
(Fig.~\ref{fig:kapferertailorshop_geodesiccycle_lengthdist}).  In the
case of the dolphin social network, the Zachary karate club network,
and the Kapferer tailor shop network, the $dk$-series $2.5k$ distribution also
shows a poor fit to the geodesic cycle length distribution when using
the exact counting method.  However in the case of the Lazega law form
network, the $dk$-series $2.5k$ distribution shows an acceptable fit to the
geodesic cycle length distribution when using the exact counting
method, while the ERGM does not.

In the Grey's Anatomy network, the exact counting method shows the
same results discussed in \citet{stivala20a}: both the ERGM and $2.5k$
distributions fit the maximum geodesic cycle length well, but the ERGM
also fits the geodesic cycle length distribution well, while the $2.5k$
distribution does not
(Fig.~\ref{fig:greysanatomy_geodesiccycle_lengthdist}). As noted in
\citet[p.~53]{stivala20a}, this is because the ERGM can reproduce the
lack of odd-length cycles in this network by using node attribute
information (most relationships in this network are heterosexual),
while the $dk$-series distributions cannot. It is, however, stated in
\citet[p.~60]{stivala20a} that of the seven four-cycles in this
network, of which all seven are also chordless, only five are
also geodesic.  This is not correct: exact counting of geodesic
cycles show that all seven are geodesic
(Fig.~\ref{fig:greysanatomy_geodesiccycle_lengthdist}).

Just as discussed in \citet{stivala20a}, in the high school friendship
network
(Fig.~\ref{fig:highschoolfriendship_geodesiccycle_lengthdist}), the
ERGM fits the maximum geodesic cycle length distribution better than
the $2.5k$ distribution does.

\subsection{Fictional character networks}
\label{sec:fiction_results}

As discussed in \citet{stivala20a}, it can be difficult to estimate ERGM
parameters for character interaction networks, not least because of
the presence of a main character, often linked to most of the other
characters, forming a high degree ``hub'' node. As noted in
\citet[p.~60]{stivala20a}, I was unable to find good converged ERGM
for the character interaction networks for Les Mis\'{e}rables, David
Copperfield, Anna Karenina, Huckleberry Finn, and hence only
$dk$-series distributions are used here for these networks. This is also the
case for the Star Wars network.

\begin{figure}
  \centering
  \includegraphics[angle=270,width=\textwidth]{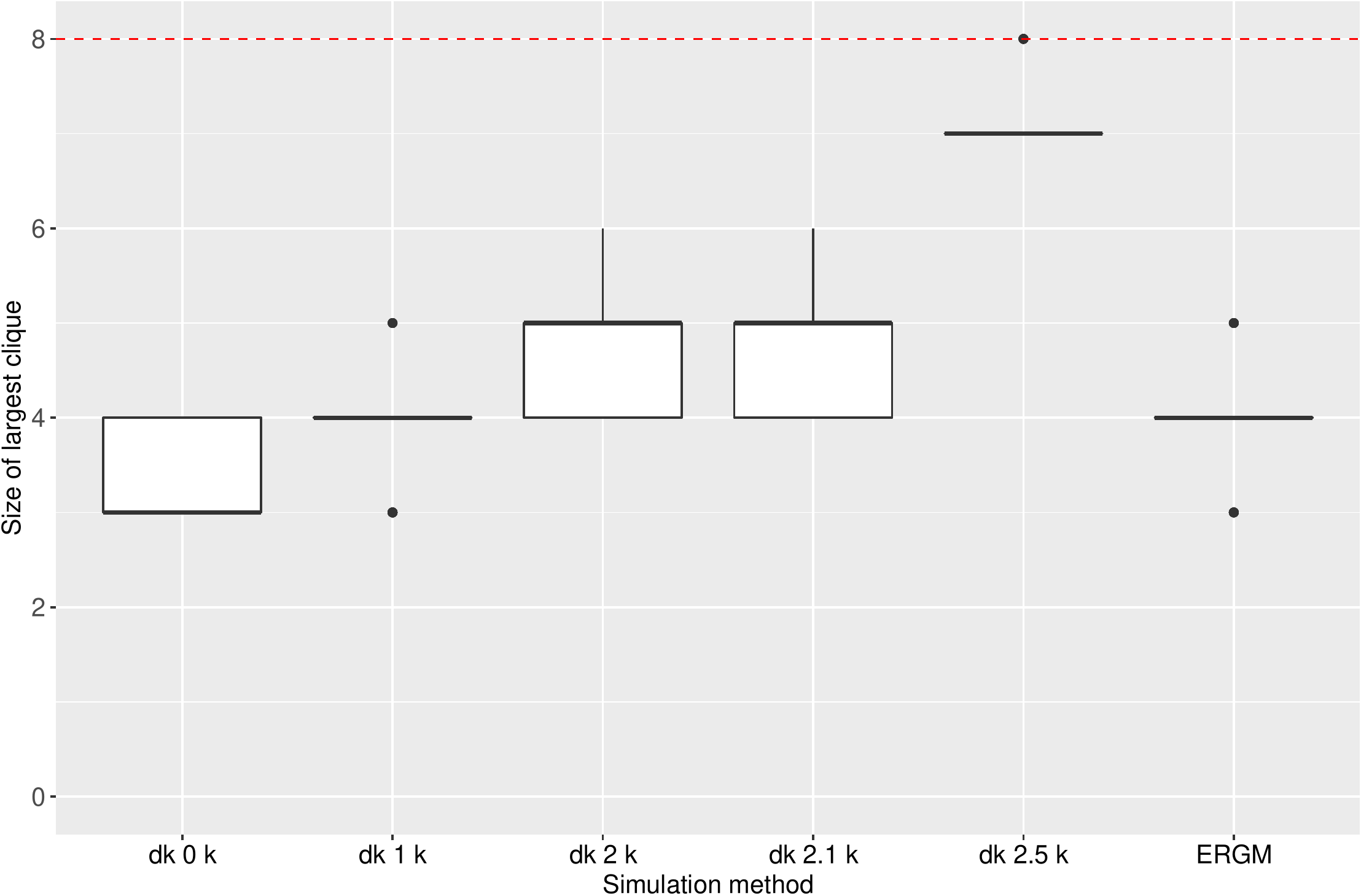}
  \caption{Largest clique sizes for the Dumbledore's Army network The
    dashed red line is the value in the observed network, with the box
    plots showing the values in 100 networks simulated from the
    $dk$-series or ERGM as labeled on the $x$-axis.  Largest clique
    sizes were computed with the \unix{igraph} R package \citep{csardi06}.}
  \label{fig:dumbledoresarmy_largest_clique_sizes}
\end{figure}

ERGM models for the two Harry Potter networks (Dumbledore's Army and
the Harry Potter peer support network) are shown, along with
goodness-of-fit plots, in \ref{sec:models}. Parameter estimates of a
simple model of the Dumbledore's Army network are shown in
Table~\ref{tab:dumbledoresarmy_ergm}. Note that, despite the
geometrically weighted edge-wise shared partners (\unix{gwesp}) term
being included in the model, the model fit to the edge-wise shared
partners distribution is not very good; in particular it fails to
reproduce the pronounced peak in the distribution at 6, or the peak in
the degree distribution at 9
(Fig.~\ref{fig:dumbledoresarmy_statnet_model1_gof}).  This could well
be because of the difficulty of an ERGM fitting the clique of size 8
clearly visible in Fig.~\ref{fig:dumbledoresarmy_network} --- a clique
that includes, as we would expect, Harry Potter himself
\citep[Fig.~2]{everton22}.
Fig.~\ref{fig:dumbledoresarmy_largest_clique_sizes} shows that, indeed,
the ERGM does not reproduce the large (8 node) clique in the observed
network; only the $2.5k$ distribution is close to doing so.

Parameter estimates for an ERGM for the Harry Potter peer support
network are shown in Table~\ref{tab:harrypotter_ergm}. I was unable to
find a converged model when the \unix{gwesp} or \unix{gwdsp} (geometrically
weighted dyad-wise shared partners) terms were included, and so this
model is a ``dyad-independent'' model \citep{koskinen13,ergm4}, as it
contains only the edge parameter, and nodal attribute parameters
relating to a single node (\unix{nodefactor}) or dyad (\unix{nodematch}).
Unsurprisingly, given the lack of a model term (such as \unix{gwesp})
to account for transitivity, the goodness-of-fit plot shows
a poor fit to the edge-wise shared partner distribution
(Fig.~\ref{fig:harrypotter_statnet_model1_gof}).

In attempting to use graph statistics of character networks for book
genre classification, \citet{holanda19} find that all of the networks
they examine are disassortative, meaning that characters with high
degree interact preferentially with those of low degree. They
therefore concluded that assortativity could not distinguish these
genres. The works they examined comprised three genres: biographical,
legendary, and fiction (the four books they classified as fiction
included two of those examined in this work: Huckleberry Finn and David
Copperfield). Here we find, similarly, that six of the seven networks
(all of which are fiction) are disassortative
(Table~\ref{tab:graph_stats_summary}). The exception, with positive
assortativity coefficient, is the Dumbledore's Army network from the
Harry Potter books. This network is notably different from the others
in that it is neither a chapter co-occurrence nor character
interaction (or narrative peer support) network, but a network
representing trust or mutual understanding, between only a subset of
the characters. Specifically it is an informal ``dark network''
\citep{cunningham16} of a resistance movement against the Death Eaters
\citep{everton22}. The relatively large clique in this network
contributes to it having a positive assortativity.

\begin{figure}
  \centering
  \includegraphics{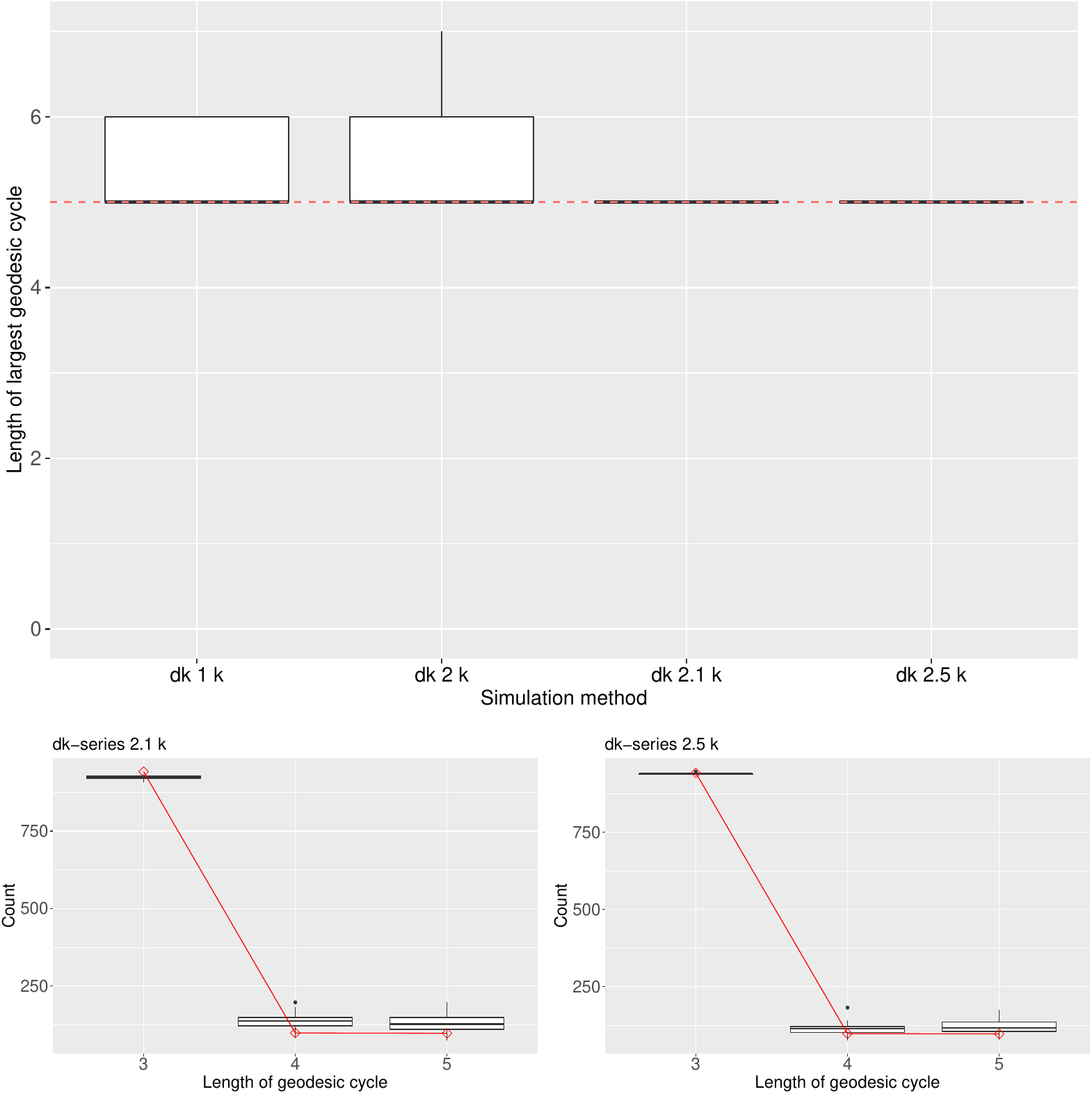}
  \caption{Largest geodesic cycle size (top), and distribution of
    geodesic cycle sizes (bottom) for the Anna Karenina network. In
    the top plot, the dashed red line is the value in the observed
    network, with the box plots showing the values in 100 networks
    simulated from the $dk$-series as labeled on the
    $x$-axis. In the bottom plots, the points shown as red diamonds
    joined by the red line are the values in the observed network,
    with the box plots showing the values in 100 networks simulated
    from the $dk$-series $2.1k$ distribution (left) and the $2.5k$
    distribution (right).}
  \label{fig:anna_geodesic_lengthdist}
\end{figure}

\begin{figure}
  \centering
  \includegraphics{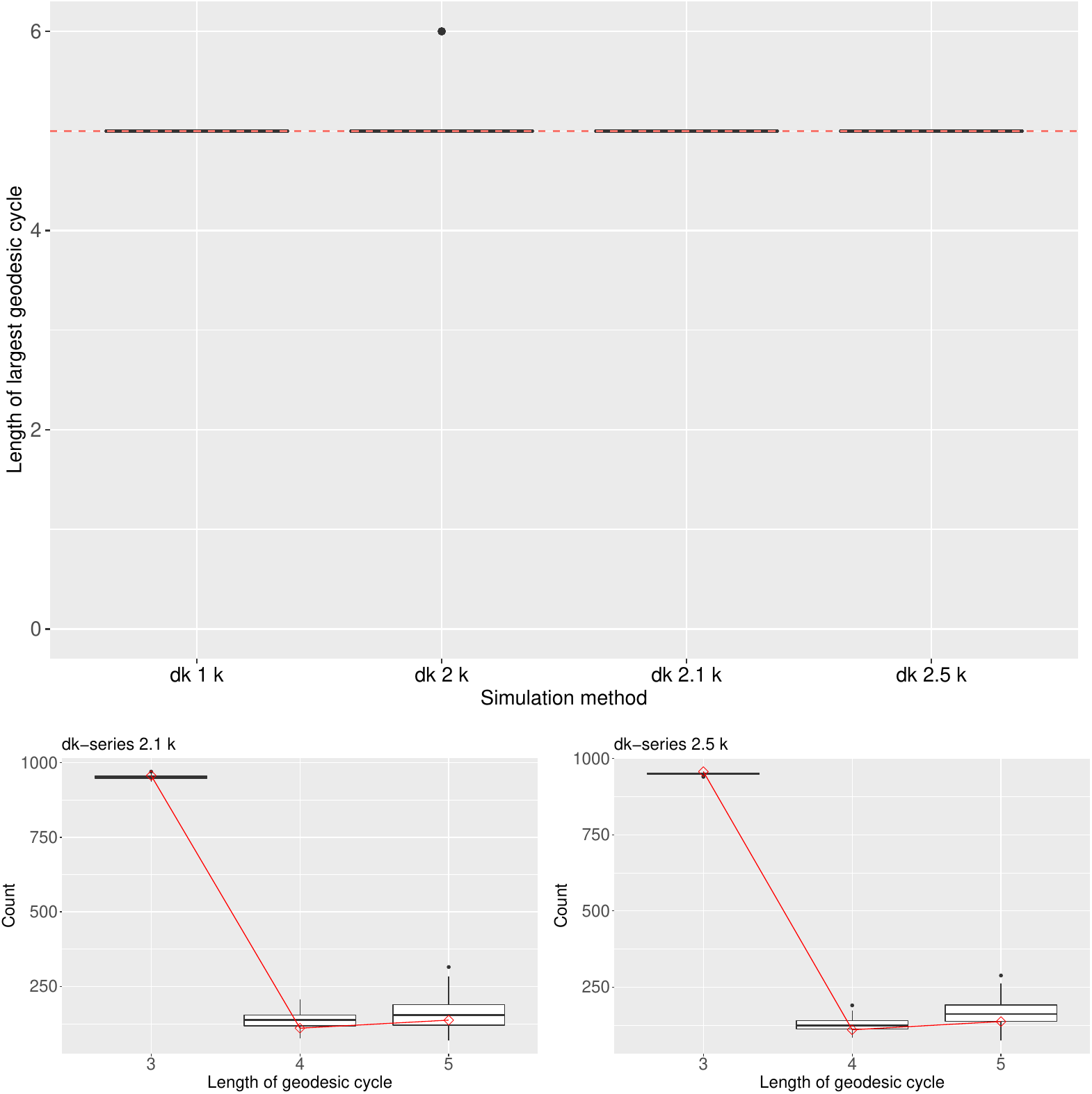}
  \caption{Largest geodesic cycle size (top), and distribution of
    geodesic cycle sizes (bottom) for the David Copperfield network. In
    the top plot, the dashed red line is the value in the observed
    network, with the box plots showing the values in 100 networks
    simulated from the $dk$-series as labeled on the
    $x$-axis. In the bottom plots, the points shown as red diamonds
    joined by the red line are the values in the observed network,
    with the box plots showing the values in 100 networks simulated
    from the $dk$-series $2.1k$ distribution (left) and the $2.5k$
    distribution (right).}
  \label{fig:david_geodesic_lengthdist}
\end{figure}

\begin{figure}
  \centering
  \includegraphics{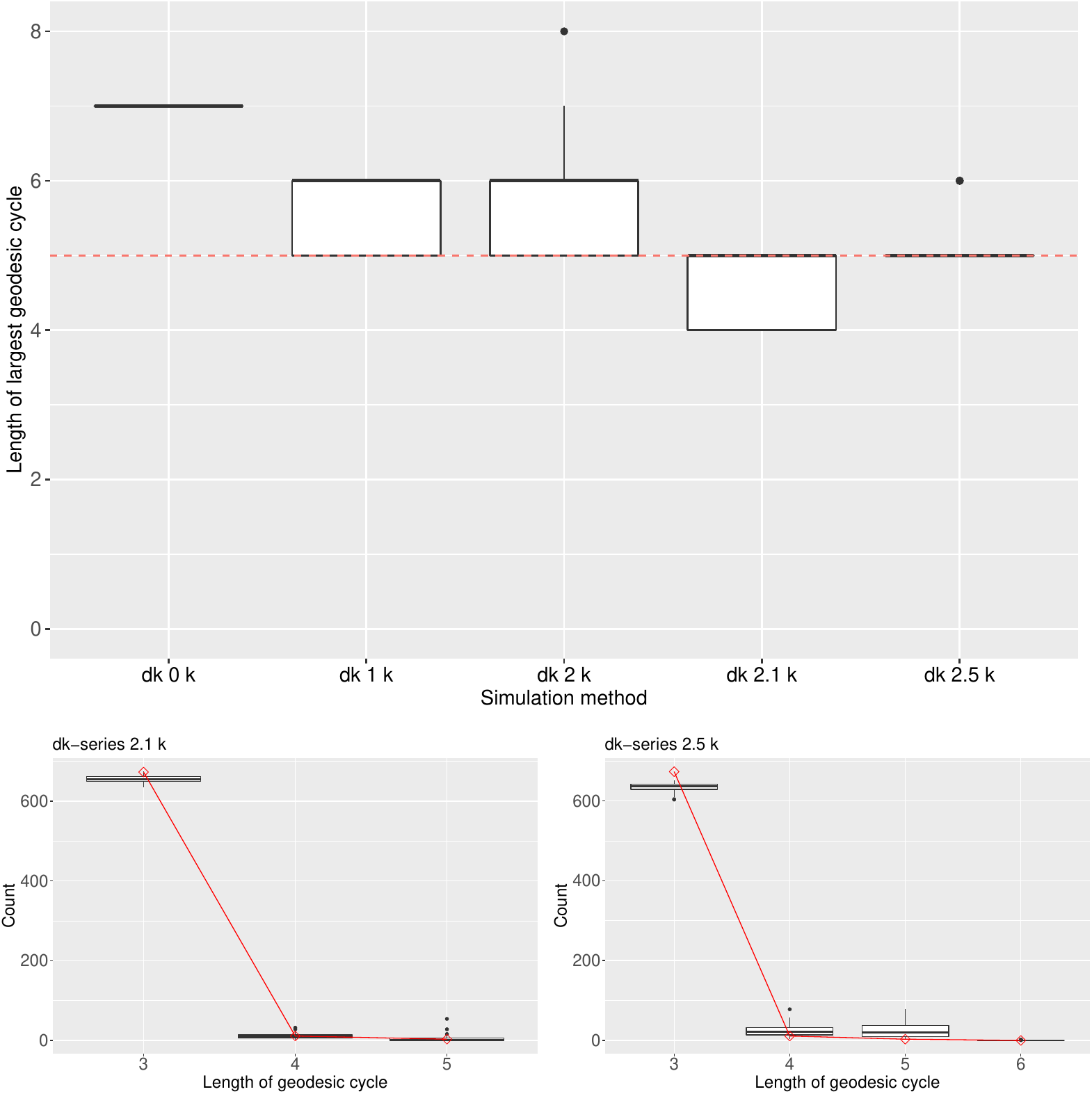}
  \caption{Largest geodesic cycle size (top), and distribution of
    geodesic cycle sizes (bottom) for the Huckleberry Finn network. In
    the top plot, the dashed red line is the value in the observed
    network, with the box plots showing the values in 100 networks
    (but only 16 networks for the $0k$ distribution)
    simulated from the $dk$-series as labeled on the
    $x$-axis. In the bottom plots, the points shown as red diamonds
    joined by the red line are the values in the observed network,
    with the box plots showing the values in 100 networks simulated
    from the $dk$-series $2.1k$ distribution (left) and the $2.5k$
    distribution (right).}
  \label{fig:huck_geodesic_lengthdist}
\end{figure}

\begin{figure}
  \centering
  \includegraphics{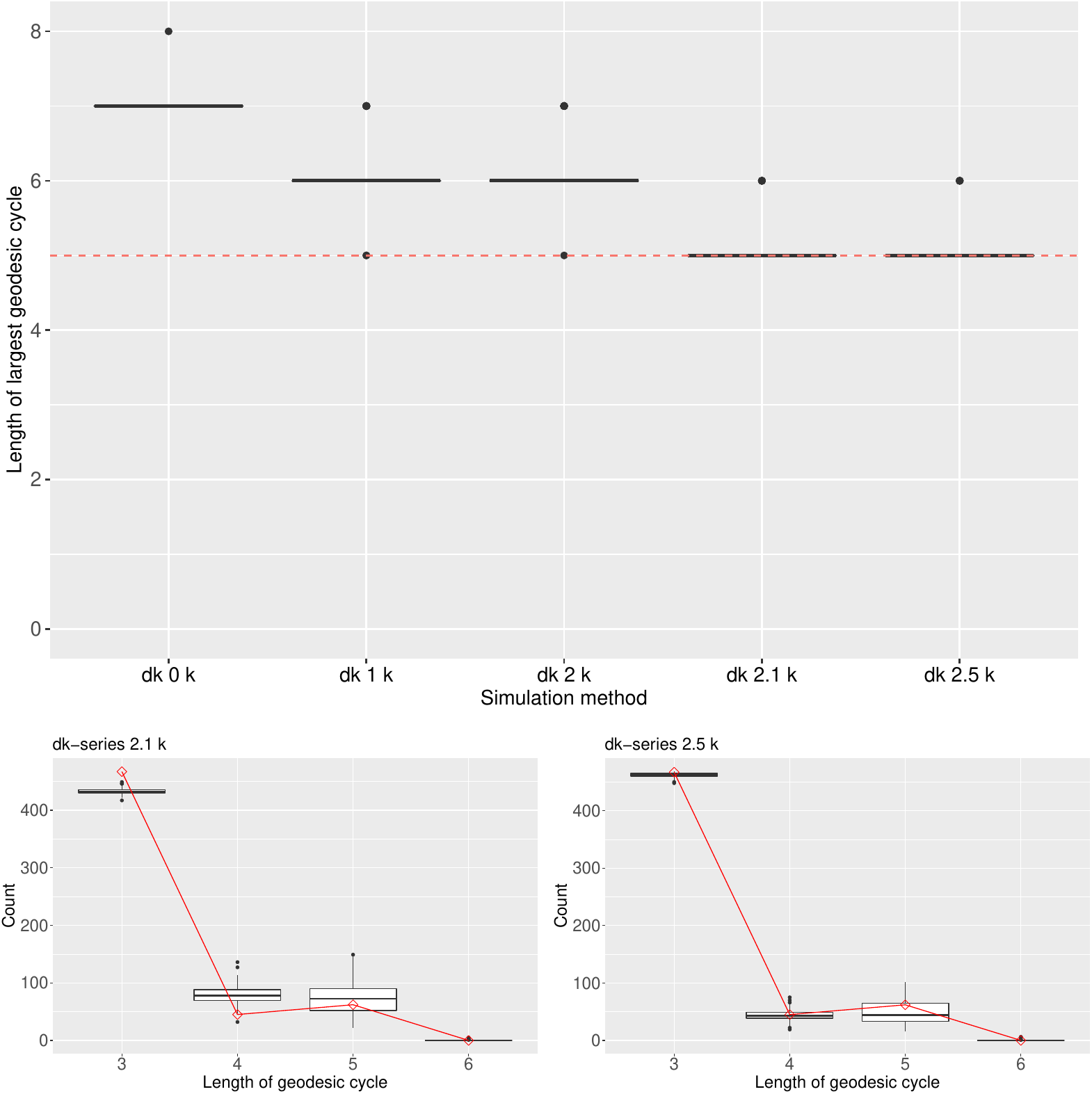}
  \caption{Largest geodesic cycle size (top), and distribution of
    geodesic cycle sizes (bottom) for the Les Mis\'{e}rables network. In
    the top plot, the dashed red line is the value in the observed
    network, with the box plots showing the values in 100 networks
    (but only 11 networks for the $0k$ distribution)
    simulated from the $dk$-series as labeled on the
    $x$-axis. In the bottom plots, the points shown as red diamonds
    joined by the red line are the values in the observed network,
    with the box plots showing the values in 100 networks simulated
    from the $dk$-series $2.1k$ distribution (left) and the $2.5k$
    distribution (right).}
  \label{fig:lesmis_geodesic_lengthdist}
\end{figure}

\begin{figure}
  \centering
  \includegraphics{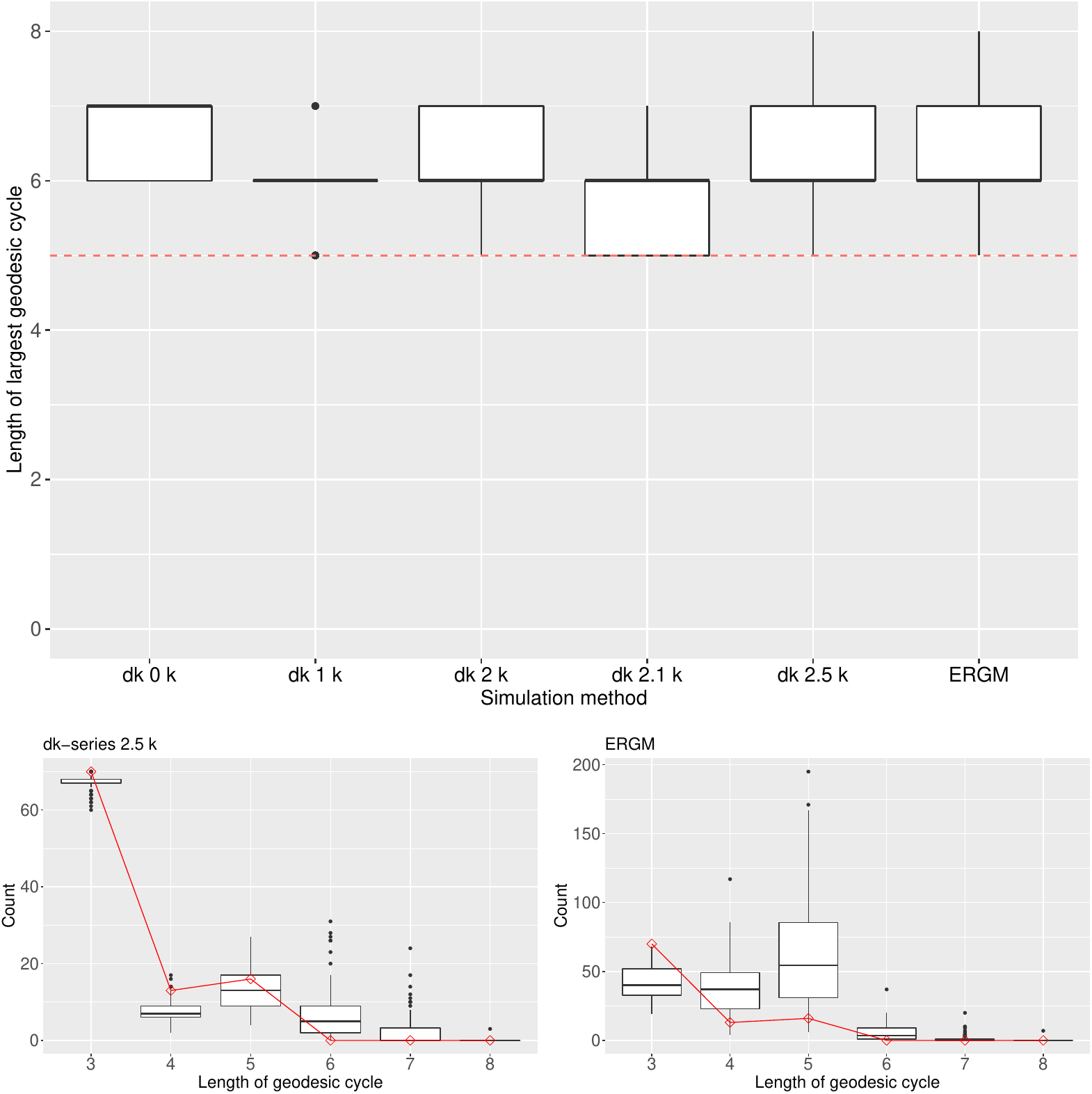}
  \caption{Largest geodesic cycle size (top), and distribution of
    geodesic cycle sizes (bottom) for the Dumbledore's Army
    network. In the top plot, the dashed red line is the value in the
    observed network, with the box plots showing the values in 100
    networks simulated from the $dk$-series or ERGM as labeled on the
    $x$-axis. In the bottom plots, the points shown as red diamonds
    joined by the red line are the values in the observed network,
    with the box plots showing the values in 100 networks simulated
    from the $dk$-series $2.5k$ distribution (left) and the ERGM
    (right).}
  \label{fig:dumbledoresarmy_geodesic_lengthdist}
\end{figure}

\begin{figure}
  \centering
  \includegraphics{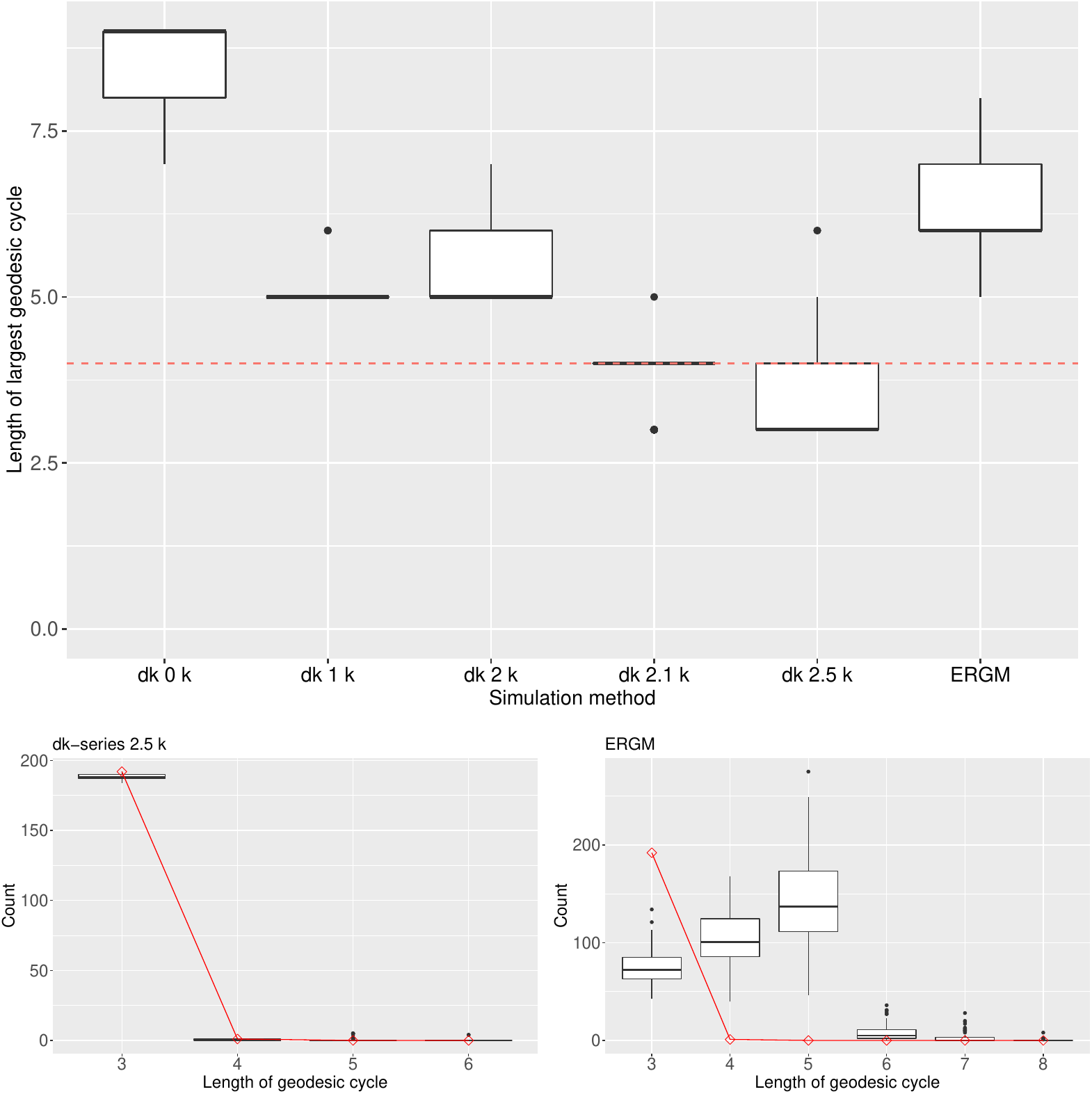}
  \caption{Largest geodesic cycle size (top), and distribution of
    geodesic cycle sizes (bottom) for the Harry Potter peer support
    network. In the top plot, the dashed red line is the value in the
    observed network, with the box plots showing the values in 100
    networks simulated from the $dk$-series or ERGM as labeled on the
    $x$-axis. In the bottom plots, the points shown as red diamonds
    joined by the red line are the values in the observed network,
    with the box plots showing the values in 100 networks simulated
    from the $dk$-series $2.5k$ distribution (left) and the ERGM
    (right).}
  \label{fig:harrypotter_geodesic_lengthdist}
\end{figure}

\begin{figure}
  \centering
  \includegraphics{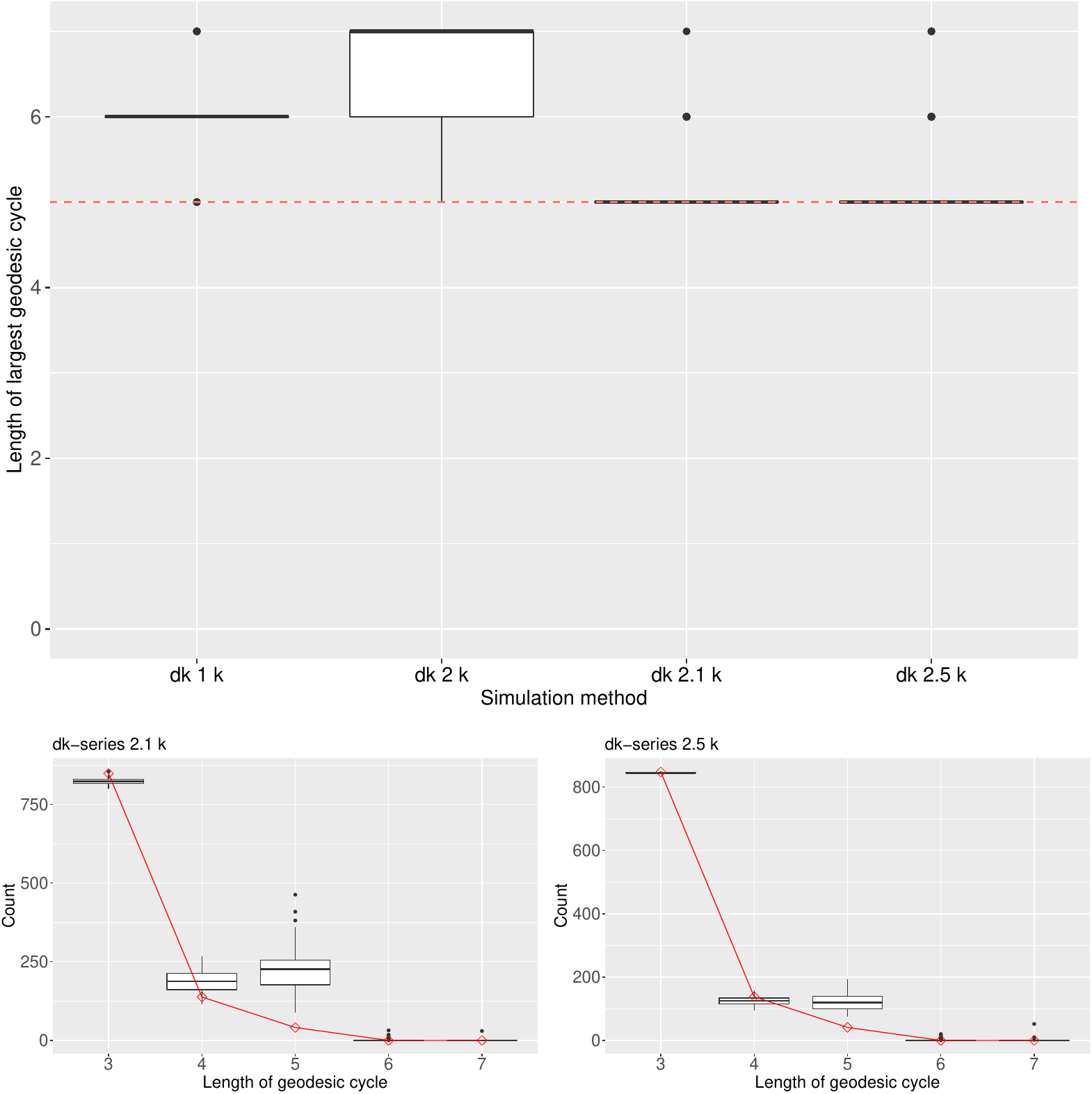}
  \caption{Largest geodesic cycle size (top), and distribution of
    geodesic cycle sizes (bottom) for the Star Wars network. In the
    top plot, the dashed red line is the value in the observed
    network, with the box plots showing the values in 100 networks
    simulated from the $dk$-series or ERGM as labeled on the
    $x$-axis. In the bottom plots, the points shown as red diamonds
    joined by the red line are the values in the observed network,
    with the box plots showing the values in 100 networks simulated
    from the $dk$-series $2.1k$ distribution (left) and the $2.5k$
    distribution (right).}
  \label{fig:starwars_geodesic_lengthdist}
\end{figure}

Figures \ref{fig:anna_geodesic_lengthdist} --
\ref{fig:starwars_geodesic_lengthdist} show the largest geodesic
cycle size and geodesic cycle size distributions for the fictional
character networks considered in this work. For the four classic
novels Anna Karenina (Fig.~\ref{fig:anna_geodesic_lengthdist}), David
Copperfield (Fig.~\ref{fig:david_geodesic_lengthdist}), Huckleberry
Finn (Fig.~\ref{fig:huck_geodesic_lengthdist}), and Les
Mis\'{e}rables (Fig.~\ref{fig:lesmis_geodesic_lengthdist}), it can be
seen that the $2.5k$ distribution fits both the largest geodesic cycle
size and the geodesic cycle size distribution well, as does the $2.1k$
distribution in most cases (it does not fit the geodesic cycle length
distribution for the Anna Karenina network,
Fig.~\ref{fig:anna_geodesic_lengthdist}, so well).
  
For the Dumbledore's Army network
(Fig.~\ref{fig:dumbledoresarmy_geodesic_lengthdist}), none of the
distributions fit the largest geodesic cycle size particularly well
(although we might consider the fit for $2.1k$ acceptable, and the
distributions are not as far from the observed value as in the
``Patricia'' 1990 and 1992 networks). The $dk$-series $2.5k$
distribution fits the observed geodesic cycle length distribution
acceptably, while the fit for the ERGM is not quite as good (having a
worse fit to geodesic cycles of length 5, for example). So the ERGM
model does not do as badly at reproducing the geodesic cycle length
distribution as it does for the largest clique size
(Fig.~\ref{fig:dumbledoresarmy_largest_clique_sizes}), relative to the
$dk$-series $2.5k$ distribution.

For the Harry Potter peer support network
(Fig.~\ref{fig:harrypotter_geodesic_lengthdist}), the $dk$-series
$2.5k$ distribution fits the geodesic cycle length distribution very
well, while the ERGM, particularly for geodesic cycles of length 5 or
less, fits it extremely poorly. The ERGM also does not reproduce the
largest geodesic cycle size, while the $2.5k$ and $2.1k$ distributions
do. As already discussed, the ERGM model for this network is a
dyad-independent model, and does not fit the edge-wise shared partner
distribution well. So in this case we can see that distributions that
reproduce transitivity, specifically $2.1k$ (average local clustering)
and $2.5k$ (clustering by degree) also reproduce the geodesic cycle
length distributions. However distributions that do not reproduce the
transitivity of the observed network, such as $0k$, $1k$, $2k$, and
the dyad-independent ERGM, also do not reproduce the geodesic cycle
length distributions.

For the Star Wars character interaction network
(Fig.~\ref{fig:starwars_geodesic_lengthdist}), both the $2.1k$ and
$2.5k$ distributions fit the maximum geodesic cycle length well, and
also fit the geodesic cycle length distribution well, with the
exception that they both produce significantly more geodesic cycles of
length 5 than are present in the observed network.

\section{Conclusions and future work}

Having compared the results in \citet{stivala20a} with the
results when counting geodesic cycles exactly instead of with the
lower bound approximation used originally, we see that the results for
the three ``Patricia'' networks are unchanged. However for some of the
other networks considered, what seemed like good fits to the geodesic
cycle length distributions for both the $dk$-series $2.5k$ and ERGM
distributions are actually not very good fits for either model when
the exact counting method is used. In one case (the Lazega law firm
network), the $2.5k$ distribution has a good fit to the geodesic cycle
length distribution while the ERGM does not. The essential conclusion
remains, however, that none of the networks examined have maximum
geodesic cycle lengths that are extremely unlikely under any of the
random graph distributions considered.

The hypothesis suggested in \citet{stivala20a}, that fictional
character networks, and specifically those from works with a single
author, might have anomalous geodesic cycle length distributions, like
the ``Patricia'' networks, is not supported on any of the seven
fictional character networks examined in this work (six of which are
from single author works). In all cases, at least one of the
$dk$-series models or ERGM reproduces the geodesic cycle length
distributions acceptably well. Hence it seems that there really is
something special about the ``Patricia'' networks in this regard,
which is not shared by fictional character networks. Just as with the
other networks considered in \citet{stivala20a}, it seems that
fictional character networks, even those from works by a single
author, have a macro-level structure, the geodesic cycle length
distribution, which can arise naturally from micro-level structures
modeled by ERGM or fit exactly by the $dk$-series random graphs.

The method used here to count geodesic cycles exactly by reading an
enumeration of all chordless cycles and testing whether or not each
one is geodesic, is rather unsatisfactory in that it involves the
enumeration of a potentially very large number of candidate chordless
cycles, the vast majority of which are not geodesic, in most cases. It
would obviously be preferable to directly enumerate geodesic cycles.
Unlike the problem of enumerating chordless cycles, there appears to
be no published work specifically on enumerating geodesic cycles. For
example, the former is listed in an enumeration of enumeration
algorithms \citep{wasa16}, and the latter is not.  The algorithm in
\citet{lokshtanov09} for finding the length of the longest geodesic
cycle, while polynomial in the size of the input graph, does not
appear to be suitable for enumerating all geodesic cycles, and was
rather designed to prove the (rather surprising) fact that finding the
longest geodesic cycle is in $\mathcal{P}$, unlike most variants of
the longest cycle problem, such as finding the longest cycle or
longest chordless cycle, which are $\mathcal{NP}$-complete
\citep{lokshtanov09}.\footnote{As noted in \citet{catrina21}, there is
an error in part of this proof, in the case of odd-length cycles, but
the proof can be corrected using the even-length case with an
auxiliary graph construction.}  Hence the former is computationally
tractable, and the latter are not (unless $\mathcal{P} =
\mathcal{NP}$).

In the absence of a more elegant and efficient algorithm, and its
implementation, for directly enumerating geodesic cycles, the method
used here could potentially be made more efficient (if not more
elegant or satisfactory from a theoretical point of view) simply by
using a more efficient method of enumerating chordless cycles. The
\citet{uno14} algorithm for enumerating chordless cycles was used in
this work, not least because it is the only one for which a publicly
available (and efficiently implemented, in the C programming language)
implementation appears to exist. Although this algorithm has been
described as the ``most notable and elegant listing algorithm''
\citep[p.~419]{ferreira14} for this problem, at least one algorithm
with better asymptotic complexity has been described
\citep{ferreira14}.  The enumeration of enumeration algorithms
\citep{wasa16} lists in addition the algorithm of \citet{wild08},
although this has no known guaranteed performance \citep{ferreira14},
and neither does the algorithm of \citet{sokhn13}, although
\citet{dias14} claims to have an improved algorithm, with a parallel
implementation for GPU (graphics processing unit) described in
\citet{jradi15}.

Although there appear to be no published papers specifically on
enumerating geodesic cycles, one set of papers addresses this issue as
part of another problem. The work of
\citet{amaldi09,amaldi10,amaldi11}, improves on the \citet{horton87}
algorithm for finding a minimum cycle basis of a graph, by restricting
the candidates to isometric (geodesic) cycles.
\citet{amaldi09} describes ``an efficient $O(nm)$ procedure which
allows to detect a single representation of each isometric cycle
without explicitly constructing the non-isometric cycles''
\citep[p.~400]{amaldi10}. This suggests that this procedure from
\citet{amaldi09} could in fact be used to efficiently enumerate all
geodesic cycles.  The creation of a publicly available implementation
of such an algorithm could greatly increase the practicality
(particularly for large graphs) of examining geodesic cycle length
distributions of empirical and random graphs for testing hypotheses
about geodesic cycles or for use as an additional goodness-of-fit test
for random graph models.

\section*{Acknowledgements}
I used the high performance computing cluster at the Institute of
Computing, Universit\`a della Svizzera italiana, for all data
processing and computations.

\section*{Data availability statement}
All code, scripts, and data are available from
\url{https://github.com/stivalaa/geodesic_cycles}.

\newpage
\appendix
\setcounter{section}{0}
\renewcommand{\thesection}{Appendix \Alph{section}}
\setcounter{table}{0}
\renewcommand{\thetable}{\Alph{section}\arabic{table}}
\setcounter{figure}{0}
\renewcommand{\thefigure}{\Alph{section}\arabic{figure}}

\renewcommand{\theHsection}{Appendix \Alph{section}}
\renewcommand{\theHfigure}{\Alph{section}\arabic{figure}}
\renewcommand{\theHtable}{\Alph{section}\arabic{table}}

\section{Visualizations of networks}
\label{sec:visualizations}

\begin{figure}[h]
  \centering
  \includegraphics[angle=270,width=.5\textwidth]{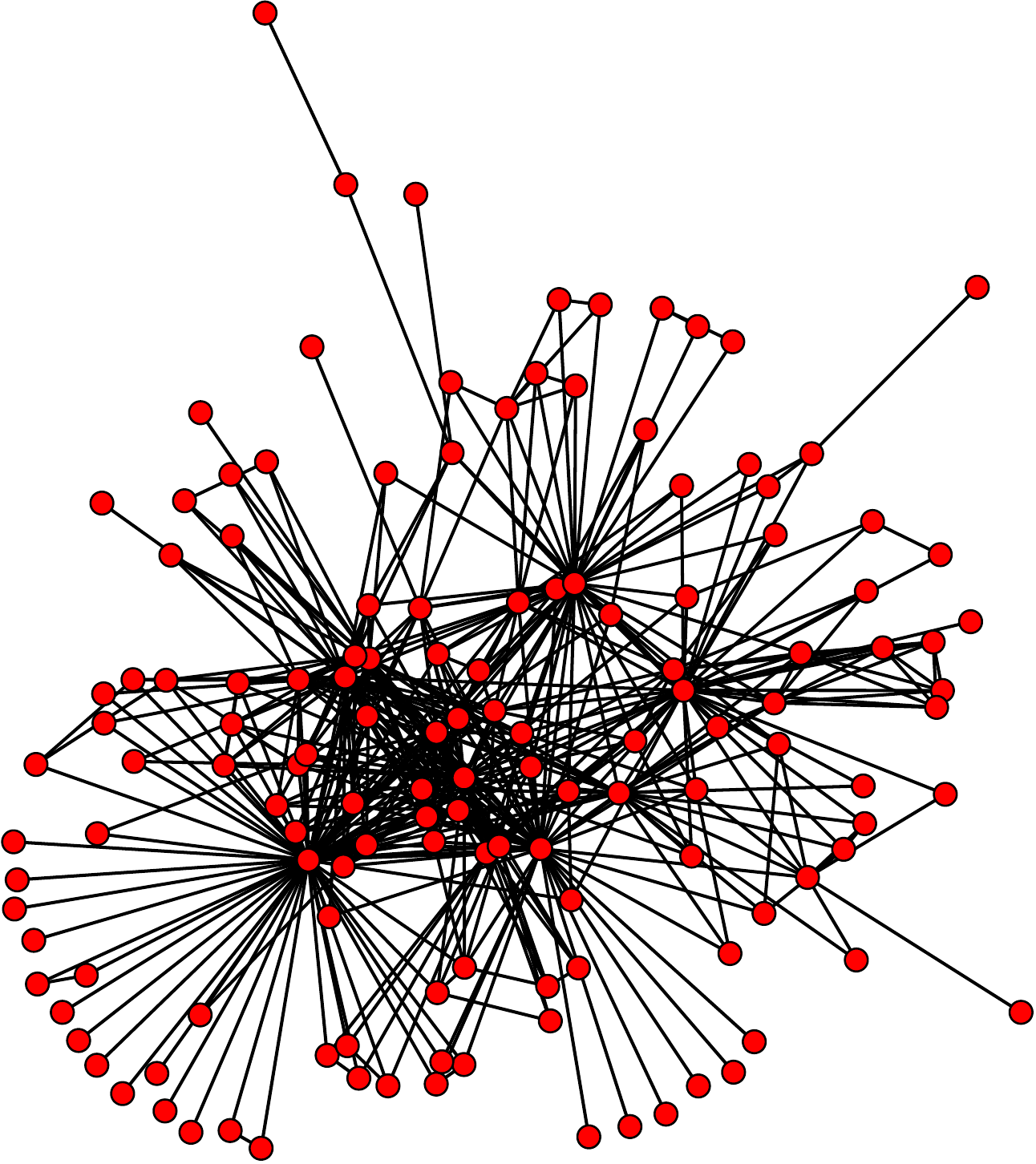}
  \caption{Anna Karenina character interaction network. Visualization
    created using the network R package \citep{butts08,network}.}
  \label{fig:annakarenina_network}
\end{figure}

\begin{figure}
  \centering
  \includegraphics[angle=270,width=.5\textwidth]{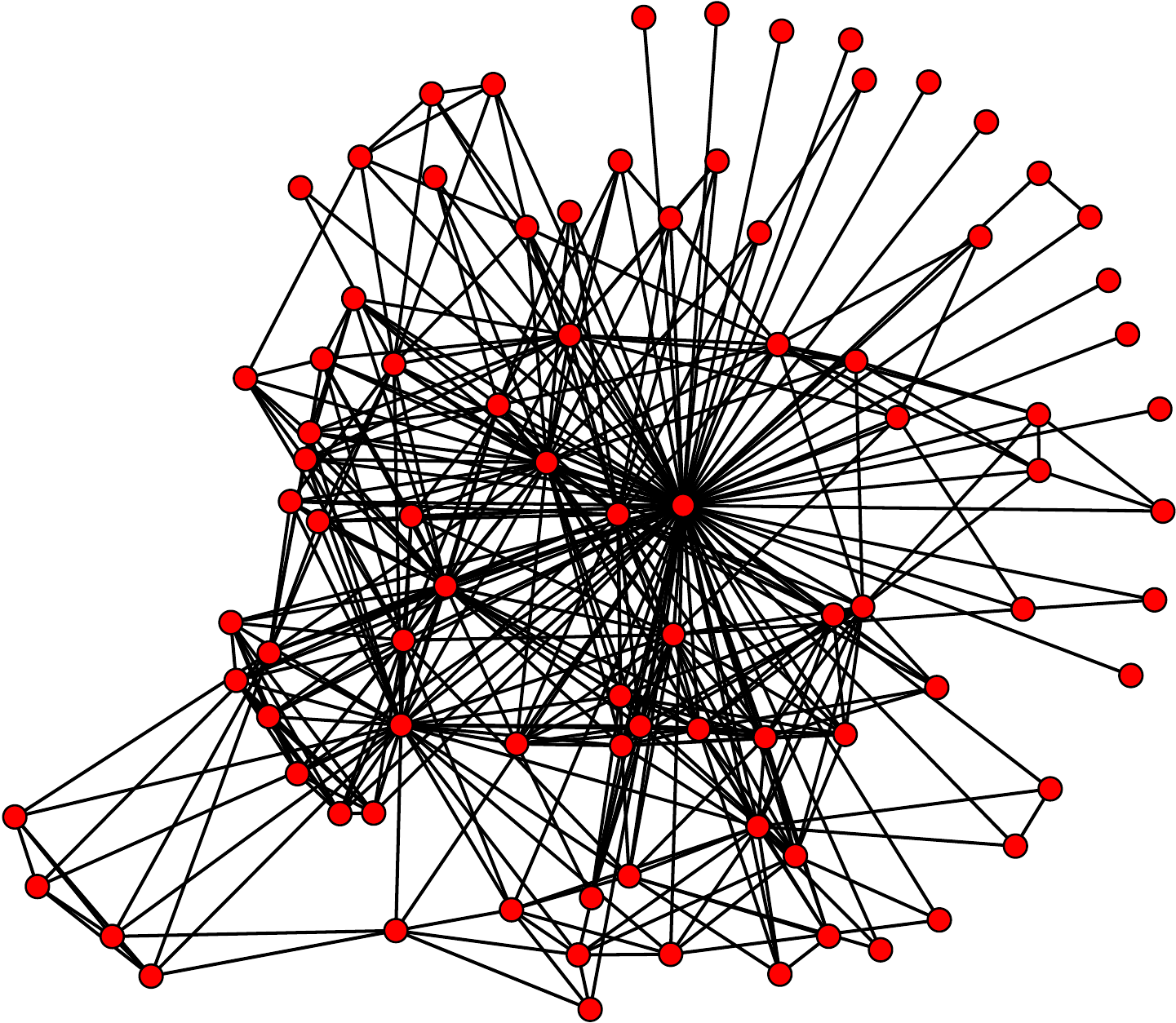}
  \caption{David Copperfield character interaction
    network. Visualization created using the network R package
    \citep{butts08,network}.}
  \label{fig:davidcopperfield_network}
\end{figure}

\begin{figure}
  \centering
  \includegraphics[angle=270,width=.5\textwidth]{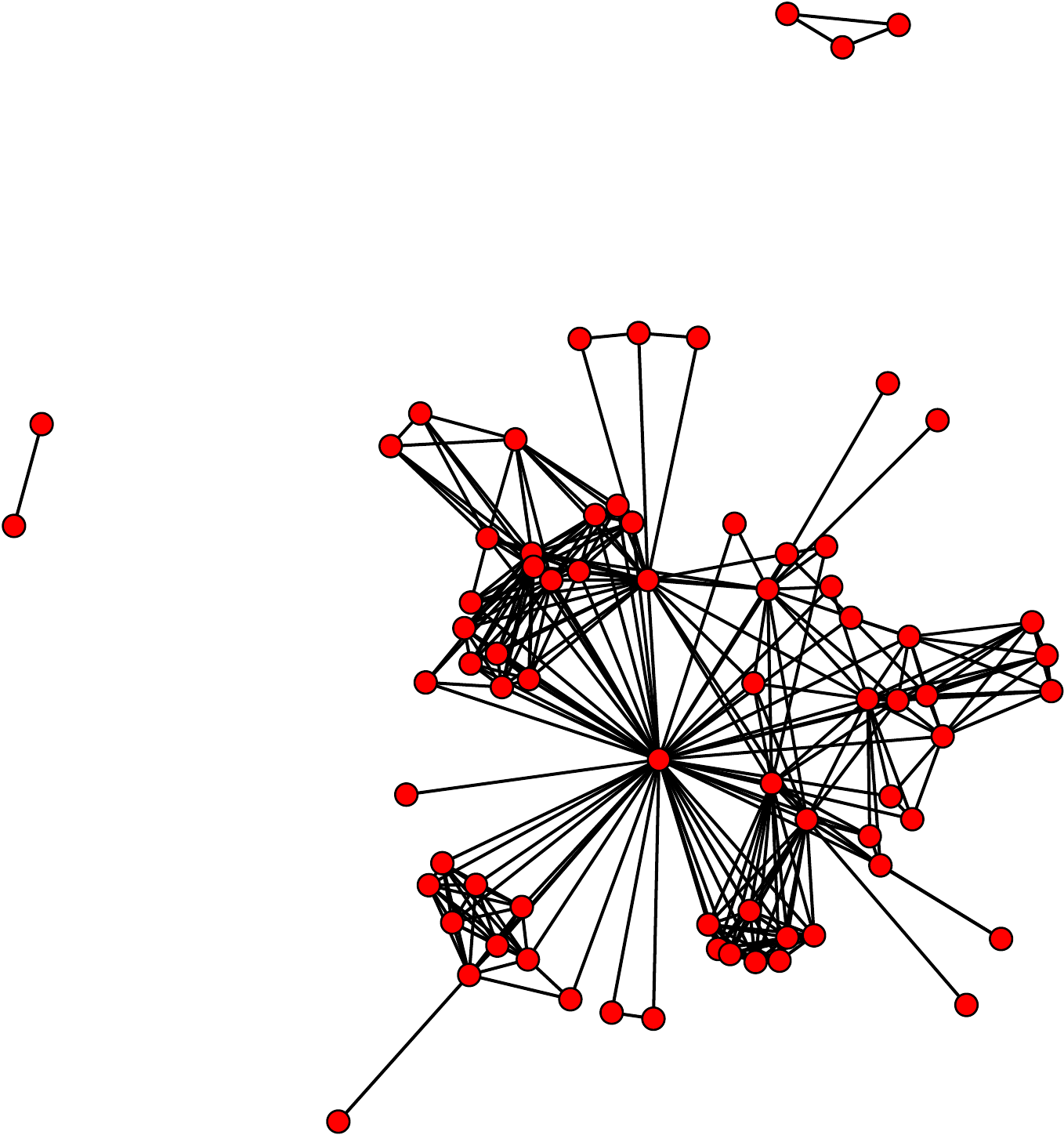}
  \caption{Huckleberry Finn character interaction
    network. Visualization created using the network R package
    \citep{butts08,network}.}
  \label{fig:huckleberryfinn_network}
\end{figure}

\begin{figure}
  \centering
  \includegraphics[angle=270,width=.5\textwidth]{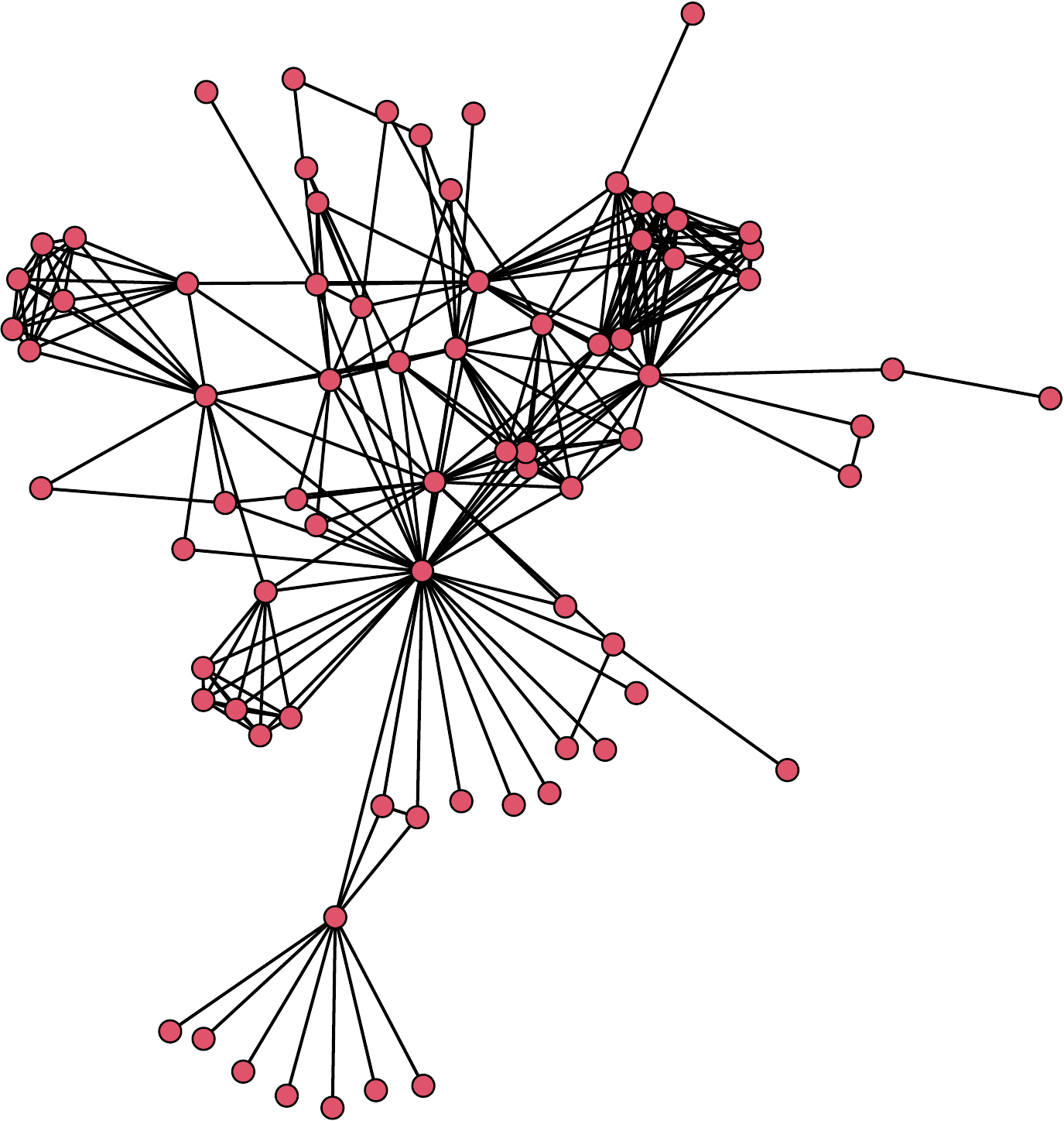}
  \caption{Les Mis\'{e}rables character interaction
    network. Visualization created using the network R package
    \citep{butts08,network}.}
  \label{fig:lesmiserables_network}
\end{figure}

\begin{figure}
  \centering
  \includegraphics[angle=270,width=.5\textwidth]{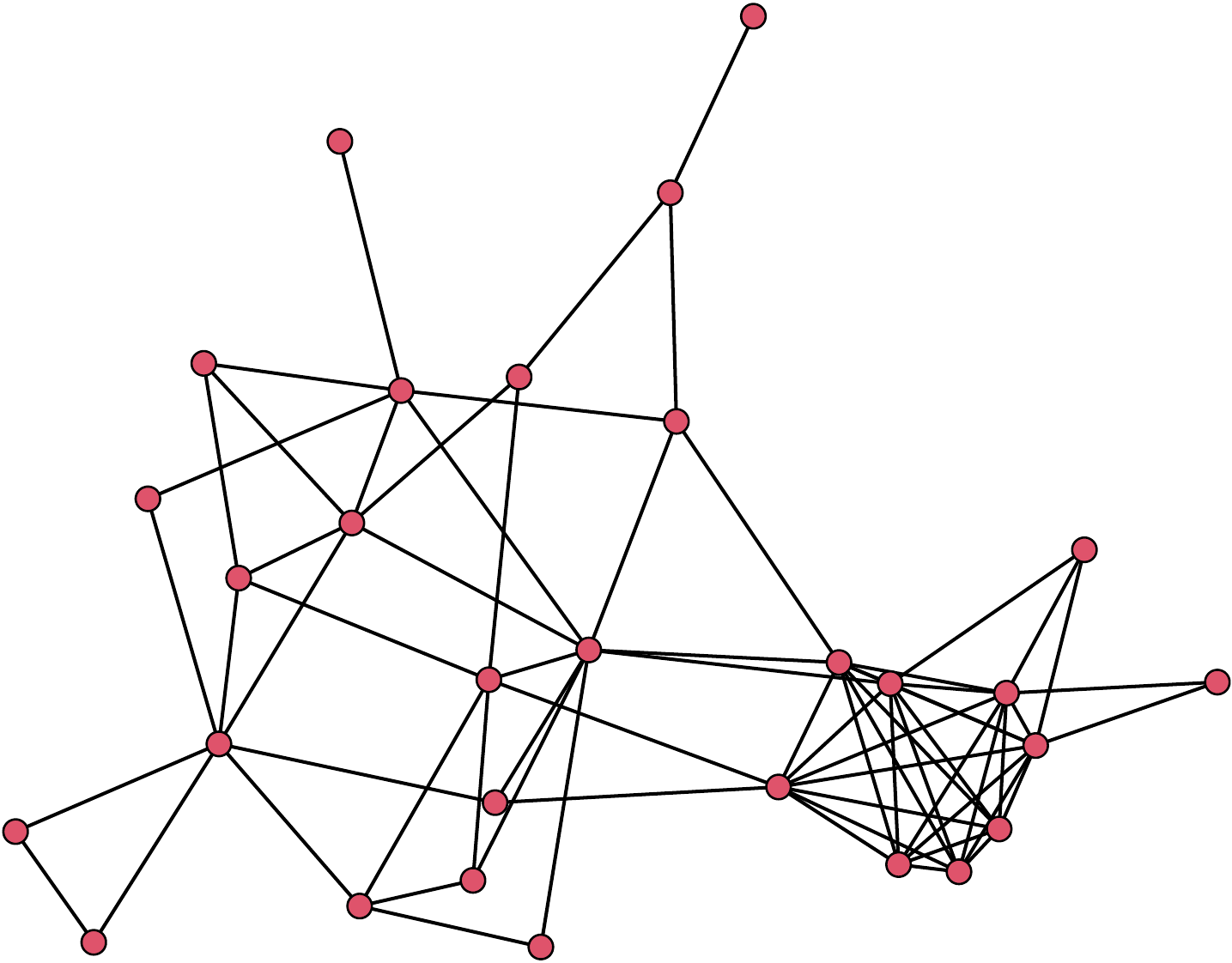}
  \caption{Dumbledore's Army
    network. Visualization created using the network R package
    \citep{butts08,network}.}
  \label{fig:dumbledoresarmy_network}
\end{figure}

\begin{figure}
  \centering
  \includegraphics[angle=270,width=.5\textwidth]{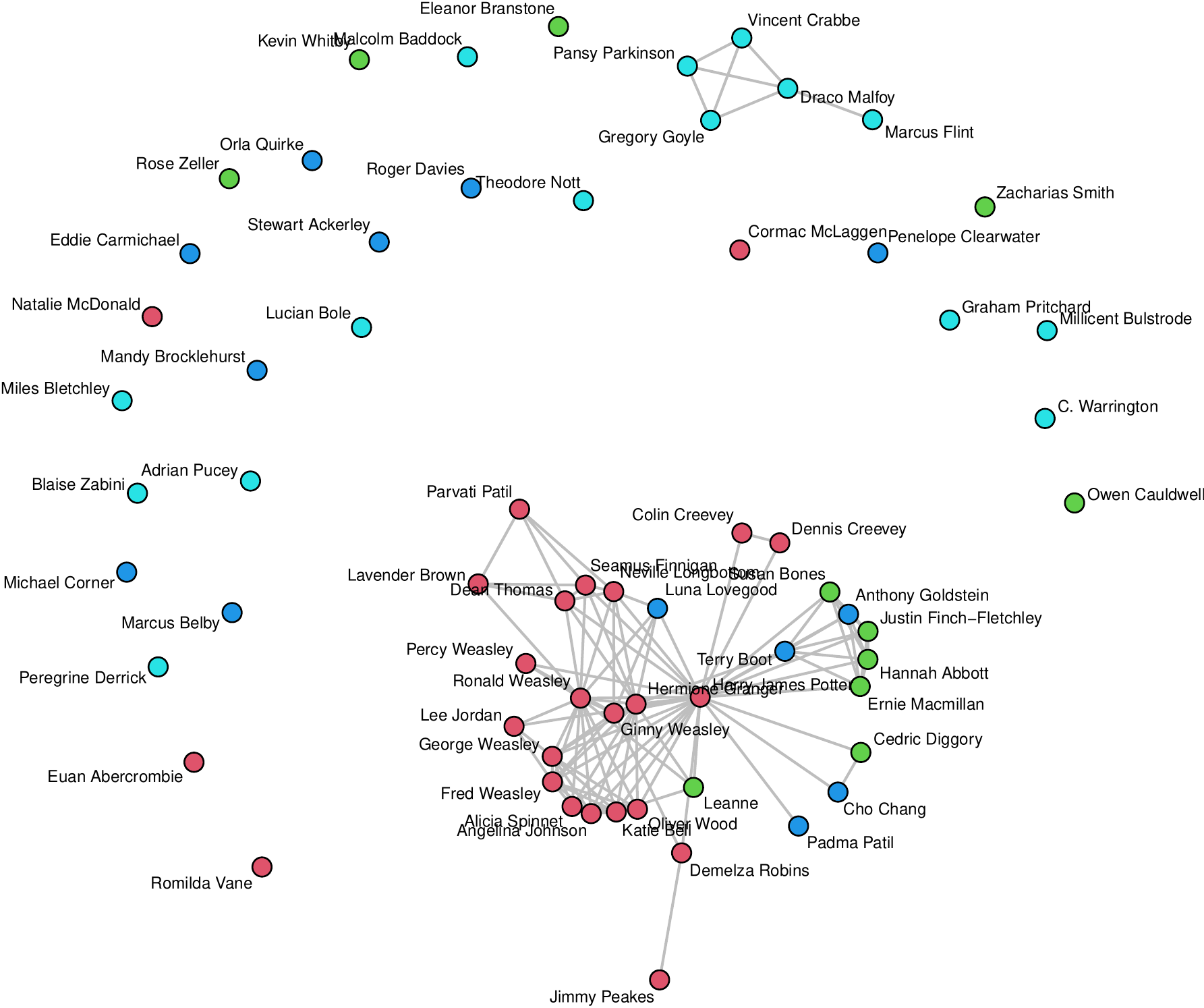}
  \caption{Harry Potter peer support network. Visualization created
    using the network R package \citep{butts08,network}. Nodes are
    coloured according to the house the character is in. }
  \label{fig:harrypotter_network}
\end{figure}

\begin{figure}
  \centering
  \includegraphics[angle=270,width=.5\textwidth]{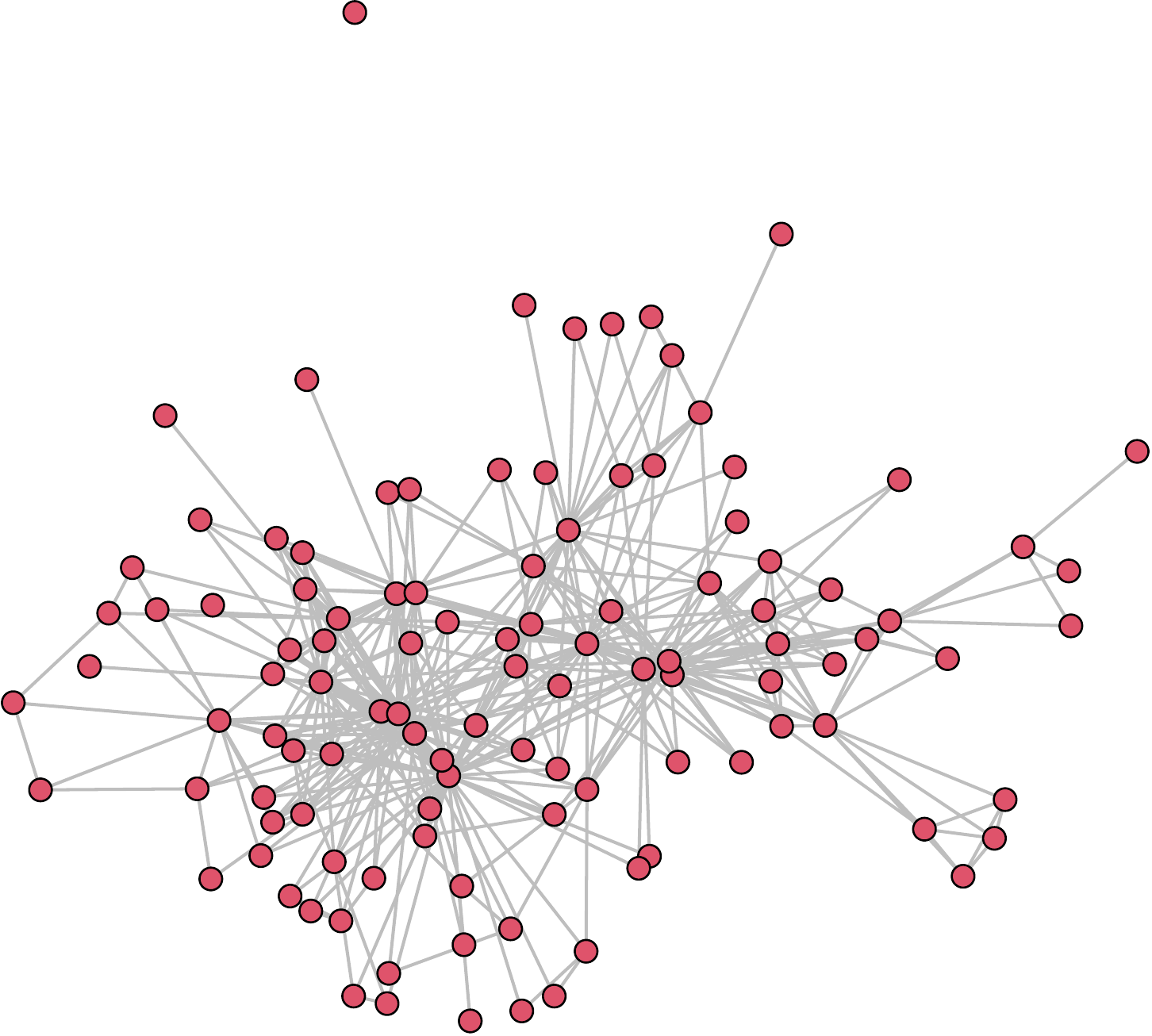}
  \caption{Star Wars character interaction network. Visualization created using
    the network R package \citep{butts08,network}.}
  \label{fig:starwars_network}
\end{figure}

\newpage
\setcounter{table}{0}
\setcounter{figure}{0}
\section{Geodesic cycle length distributions for networks from \texorpdfstring{\citet{stivala20a}}{Stivala (2020a)}}
\label{sec:reproduction}

The plots in this section reproduce, on the same plots, the results
from \citet{stivala20a}, where the geodesic cycles were counted only
approximately (lower bound) using the
\texttt{find\_large\_atomic\_cycle} algorithm \citep{gashler11,gashler12},
labeled \textsf{findAtomicCycles} on the plots, along with the
results counting geodesic cycles exactly, labeled
\textsf{countGeodesicCycles} on the plots.  In the plots of largest
geodesic cycle lengths (top plot on figures), the observed largest
geodesic cycle lengths are shown as horizontal lines, while in the
plots of geodesic cycle length distributions (bottom plots on
figures), the observed values are shown as diamonds (joined by lines
as a visual aid only). The box plots represent values from 100 (except
where otherwise noted) networks simulated from the distributions as
labeled on the plots.

\begin{figure}[h]
  \centering
  \includegraphics{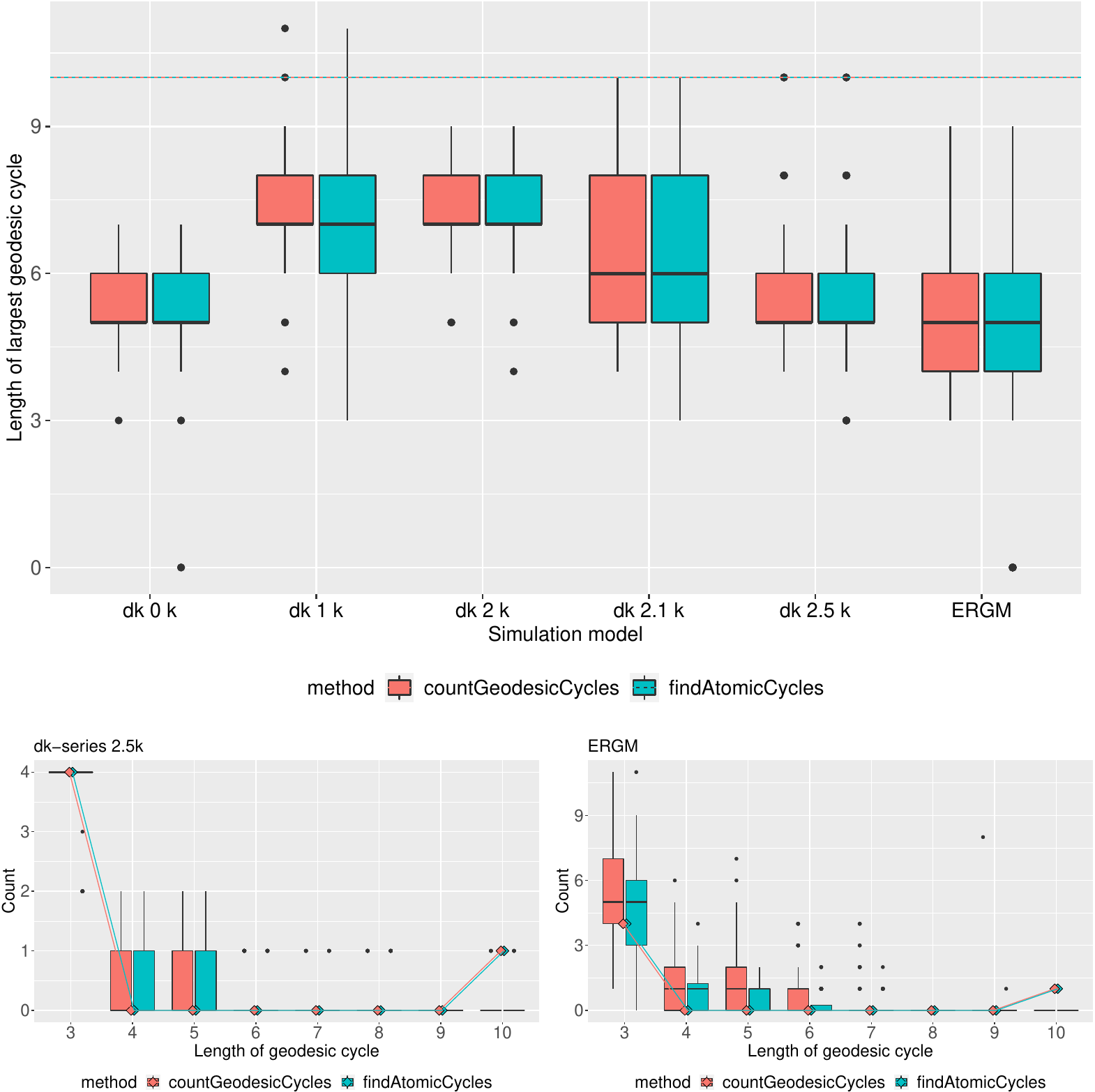}
  \caption{Largest geodesic cycle size (top), and distribution of
    geodesic cycle sizes (bottom) for Patricia's 1990 network,
    corresponding to Fig.~5 in \citet{stivala20a}.  The bottom two
    plots show results for the $dk$-series $2.5k$ distribution
    (left) and from the ERGM (right).}
  \label{fig:patricia1990_geodesiccycle_lengthdist}
\end{figure}

\begin{figure}
  \centering
  \includegraphics{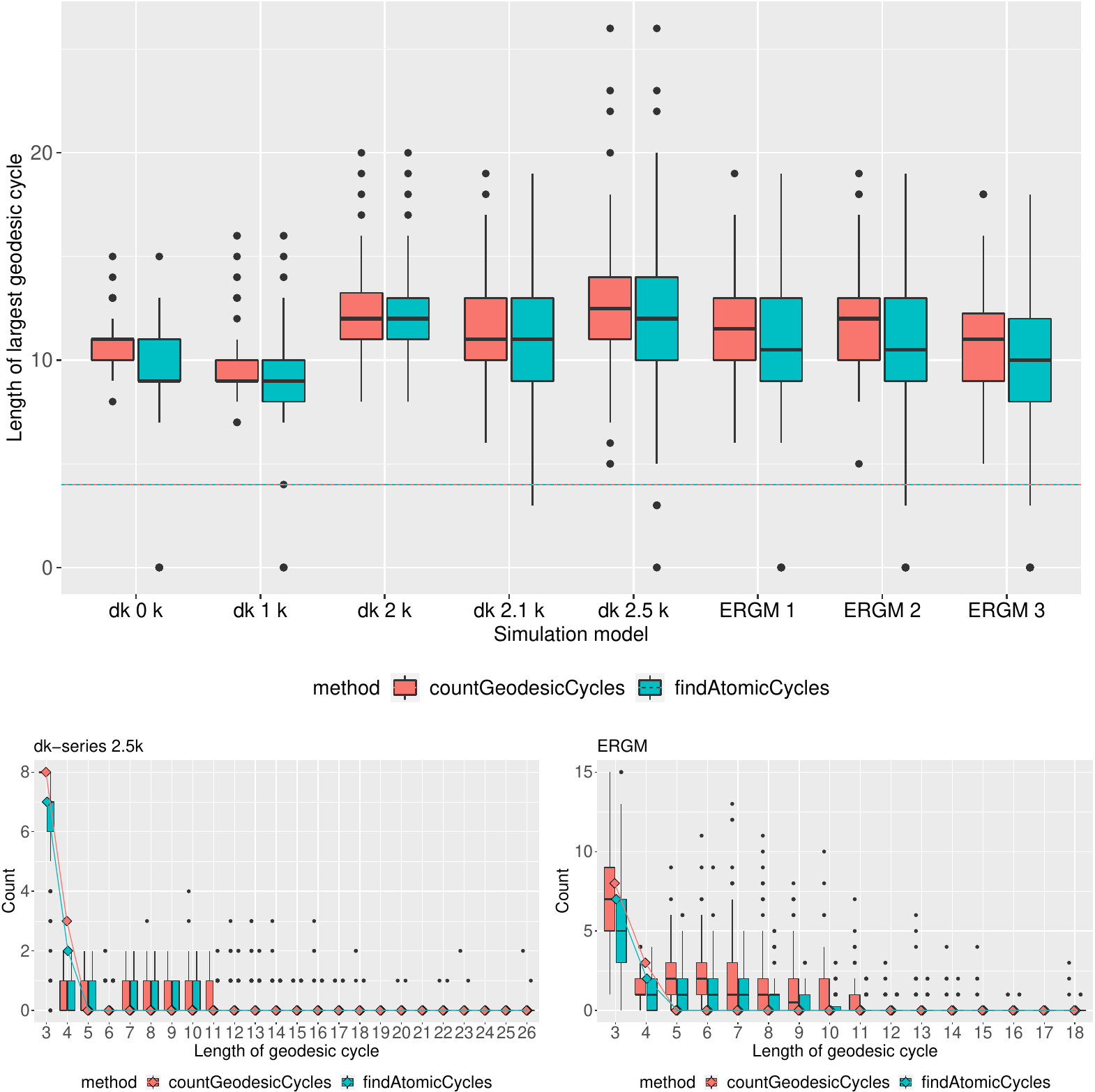}
  \caption{Largest geodesic cycle size (top), and distribution of
    geodesic cycle sizes (bottom) for Patricia's 1992 network,
    corresponding to Fig.~6 in \citet{stivala20a}. The bottom two
    plots show results for the $dk$-series $2.5k$ distribution (left)
    and from the ERGM (right).}
  \label{fig:patricia1992_geodesiccycle_lengthdist}
\end{figure}

\begin{figure}
  \centering
  \includegraphics{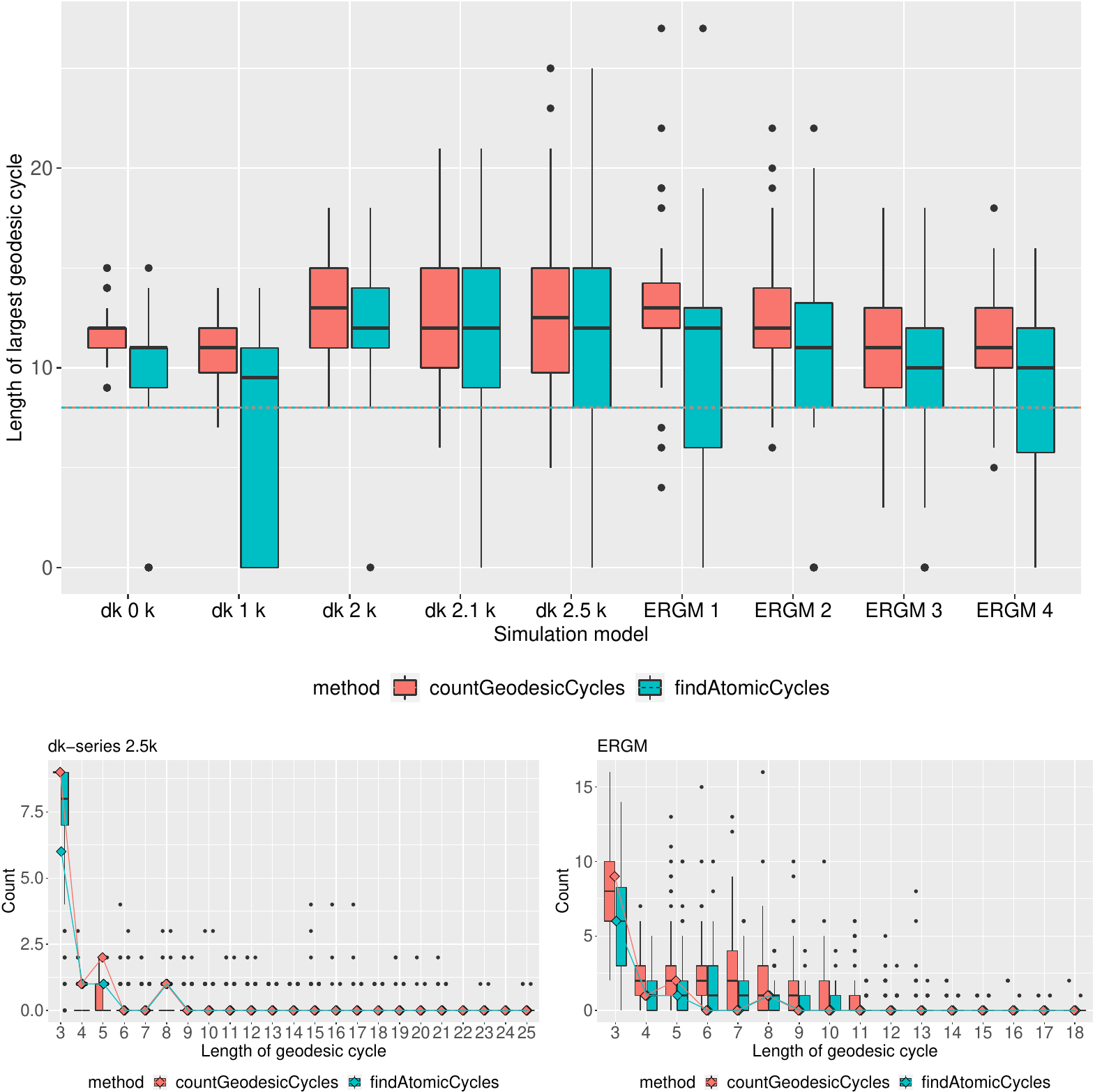}
  \caption{Largest geodesic cycle size (top), and distribution of
    geodesic cycle sizes (bottom) for Patricia's 1993 network,
    corresponding to Fig.~7 in \citet{stivala20a}.  The bottom two
    plots show results for the $dk$-series $2.5k$ distribution (left)
    and from the ERGM Model 3 (right).}
  \label{fig:patricia1993_geodesiccycle_lengthdist}
\end{figure}

\begin{figure}
  \centering
  \includegraphics{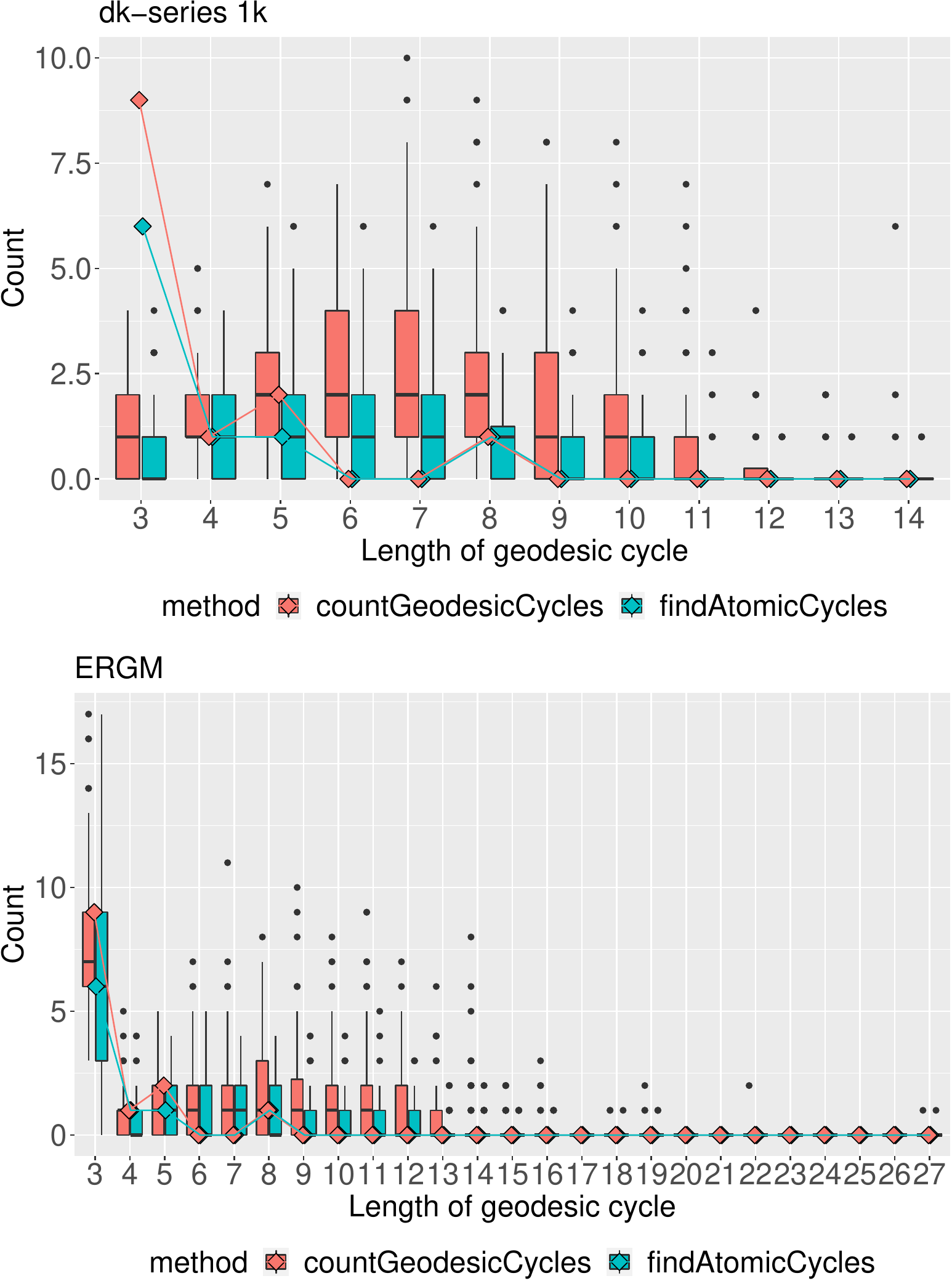}
  \caption{Distribution of geodesic cycle sizes for Patricia's 1993
    network, corresponding to Fig.~8 in \citet{stivala20a}. These
    results are for the $dk$-series $1k$ distribution (top) and ERGM
    Model 1 (bottom).}
  \label{fig:patricia1993_geodesiccycle_more_lengthdist}
\end{figure}

\begin{figure}
  \centering
  \includegraphics{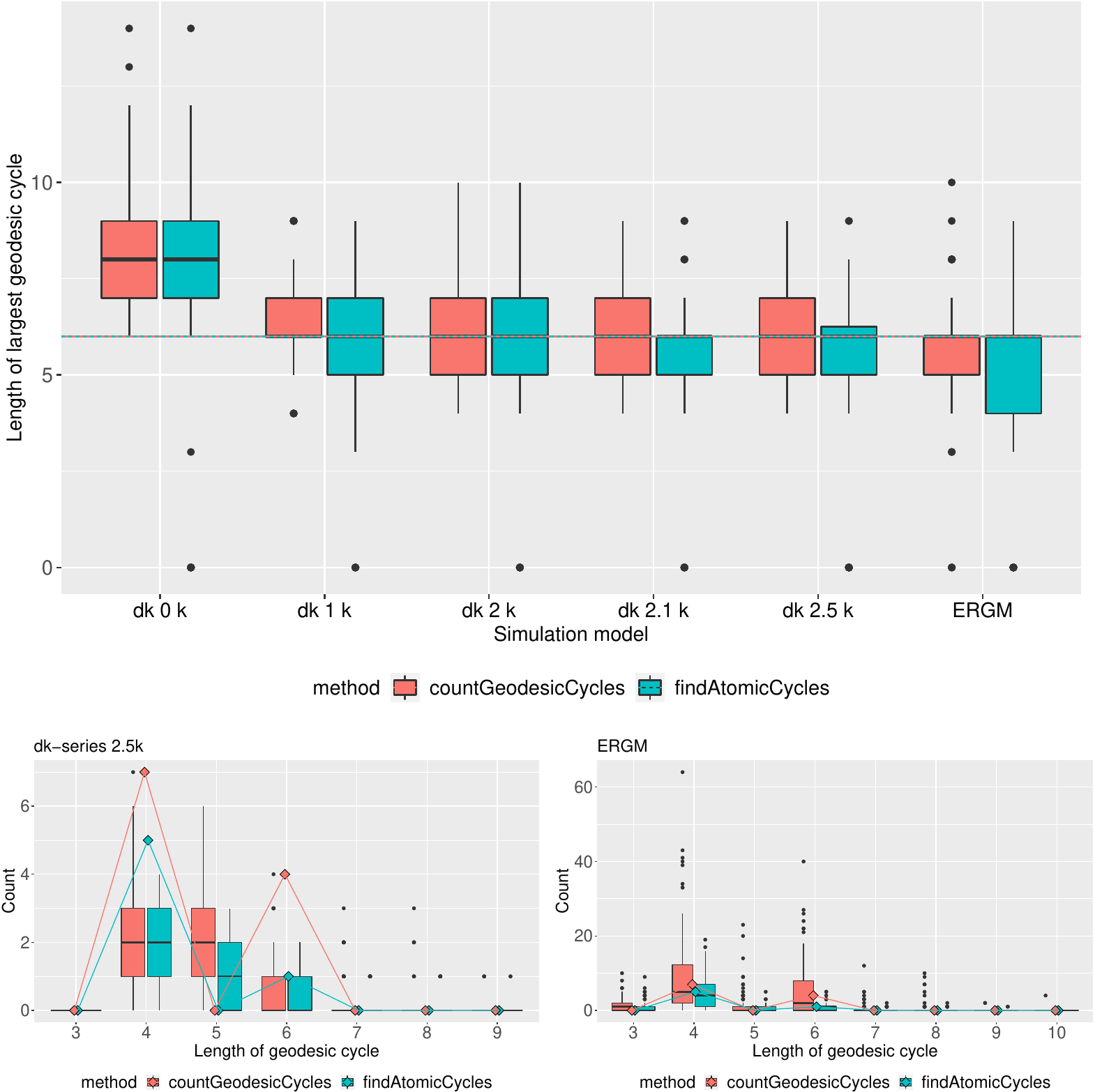}
  \caption{Largest geodesic cycle size (top), and distribution of
    geodesic cycle sizes (bottom) for the Grey's Anatomy sexual
    contact network, corresponding to Fig.~9 in \citet{stivala20a}.
    The bottom two plots show results
    for the $dk$-series $2.5k$ distribution (left) and from the ERGM
    (right).}
  \label{fig:greysanatomy_geodesiccycle_lengthdist}
\end{figure}

\begin{figure}
  \centering
  \includegraphics{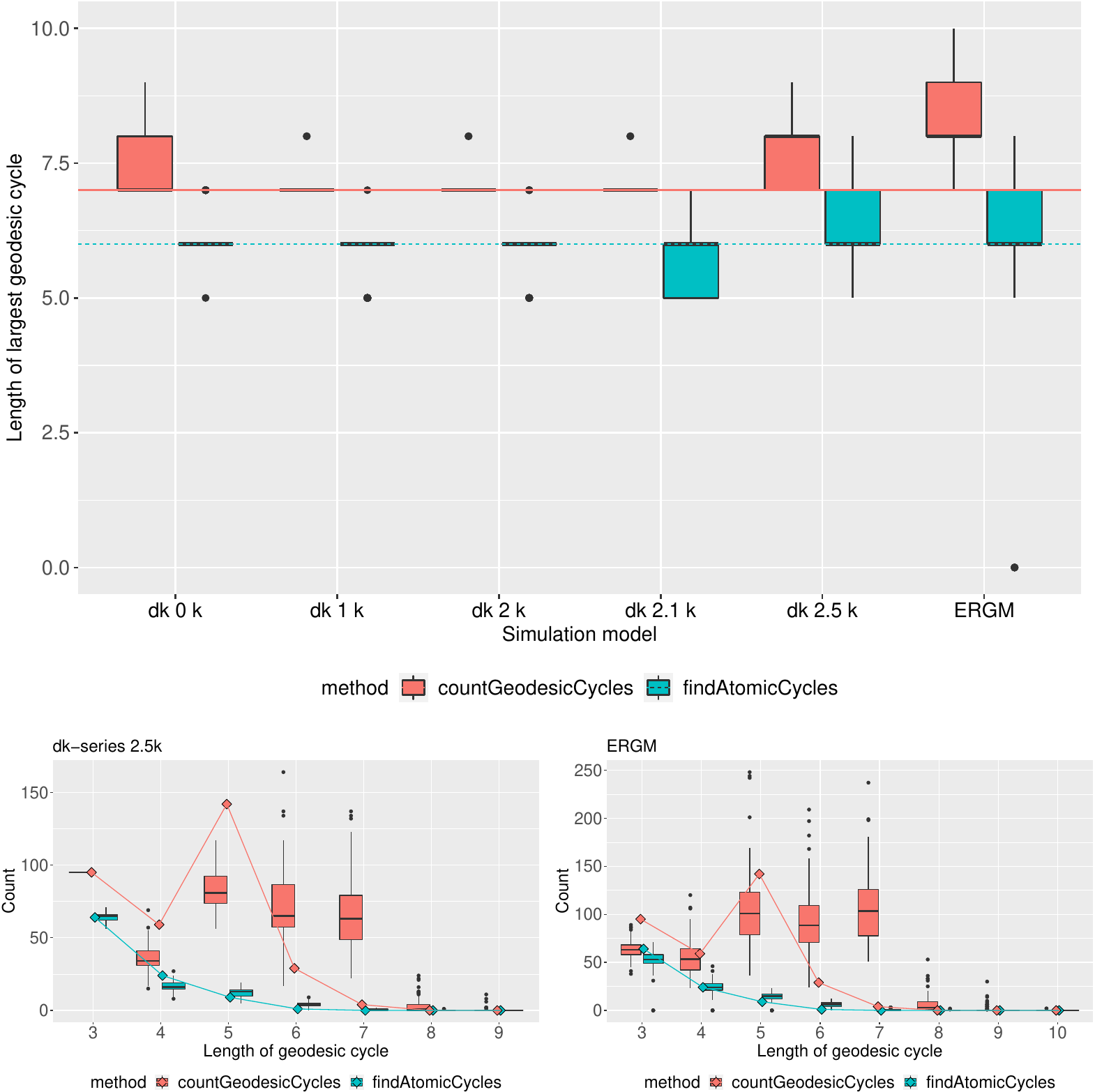}
  \caption{Largest geodesic cycle size (top), and distribution of
    geodesic cycle sizes (bottom) for the dolphin social network,
    corresponding to Fig.~10 in \citet{stivala20a}. The bottom two
    plots show results for the $dk$-series $2.5k$ distribution (left)
    and from the ERGM (right).}
  \label{fig:dolphins_geodesiccycle_lengthdist}
\end{figure}

\begin{figure}
  \centering
  \includegraphics{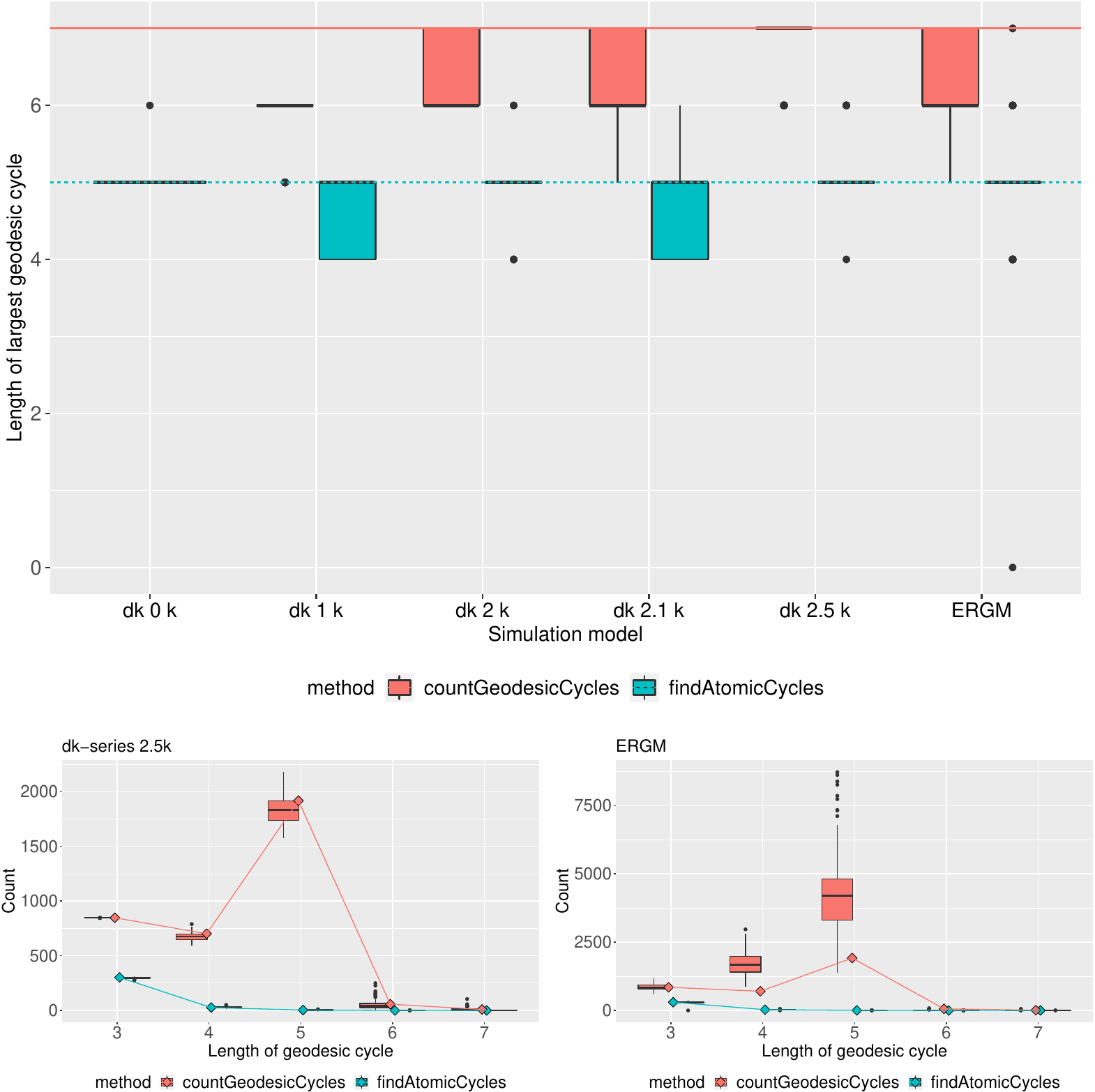}
  \caption{Largest geodesic cycle size (top), and distribution of
    geodesic cycle sizes (bottom) for the Lazega law firm friendship
    network, corresponding to Fig.~11 in \citet{stivala20a}.
    The bottom two plots show results for the $dk$-series
    $2.5k$ distribution (left) and from the ERGM (right).}
    \label{fig:lawfirm_geodesiccycle_lengthdist}
 \end{figure}

\begin{figure}
  \centering
  \includegraphics{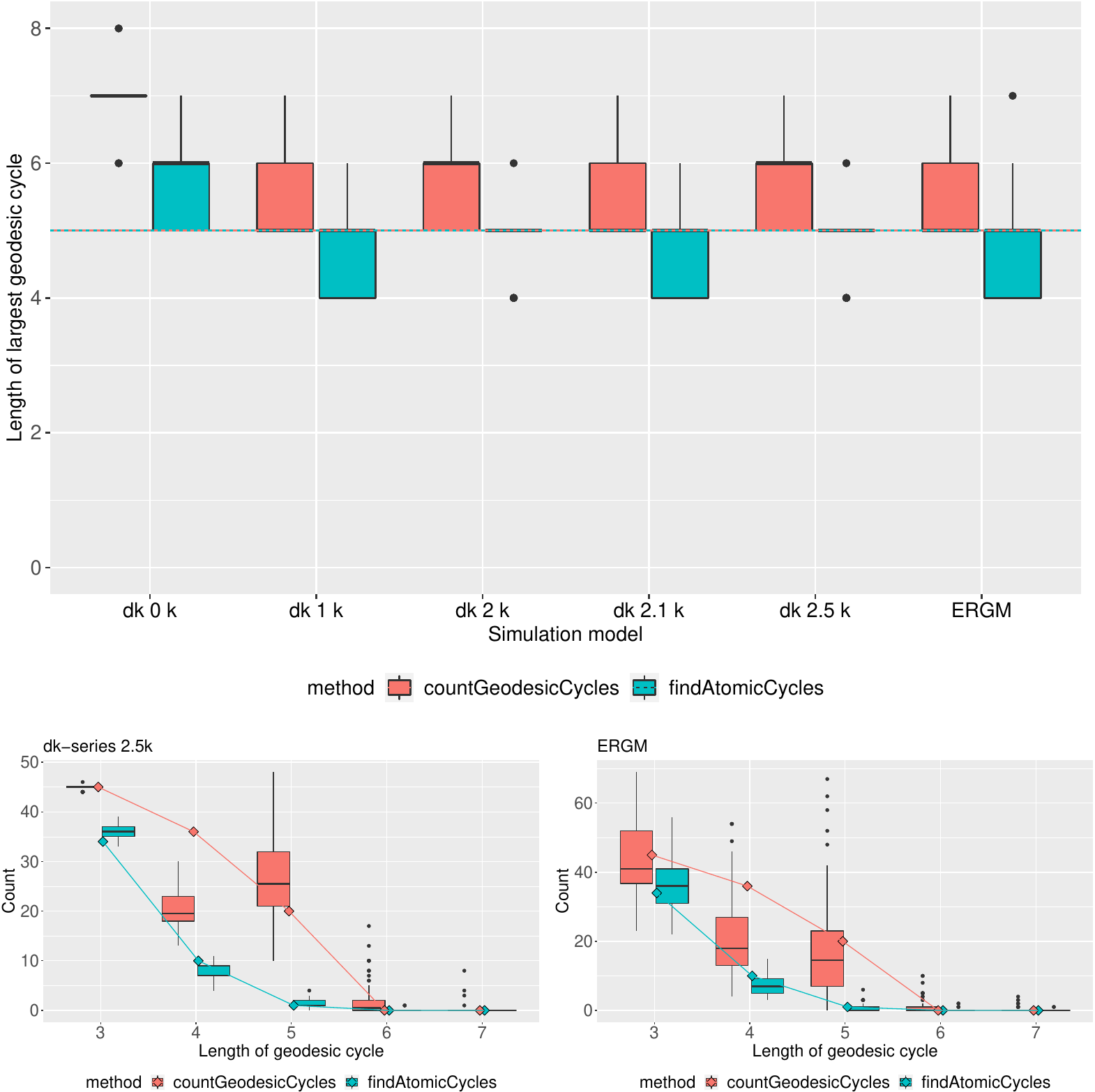}
  \caption{Largest geodesic cycle size (top), and distribution of
    geodesic cycle sizes (bottom) for the Zachary karate club network,
    corresponding to Fig.~12 in \citet{stivala20a}.  The bottom two
    plots show results for the $dk$-series $2.5k$ distribution (left)
    and from the ERGM (right).}
  \label{fig:zacharykarateclub_geodesiccycle_lengthdist}
\end{figure}

\begin{figure}
  \centering
  \includegraphics{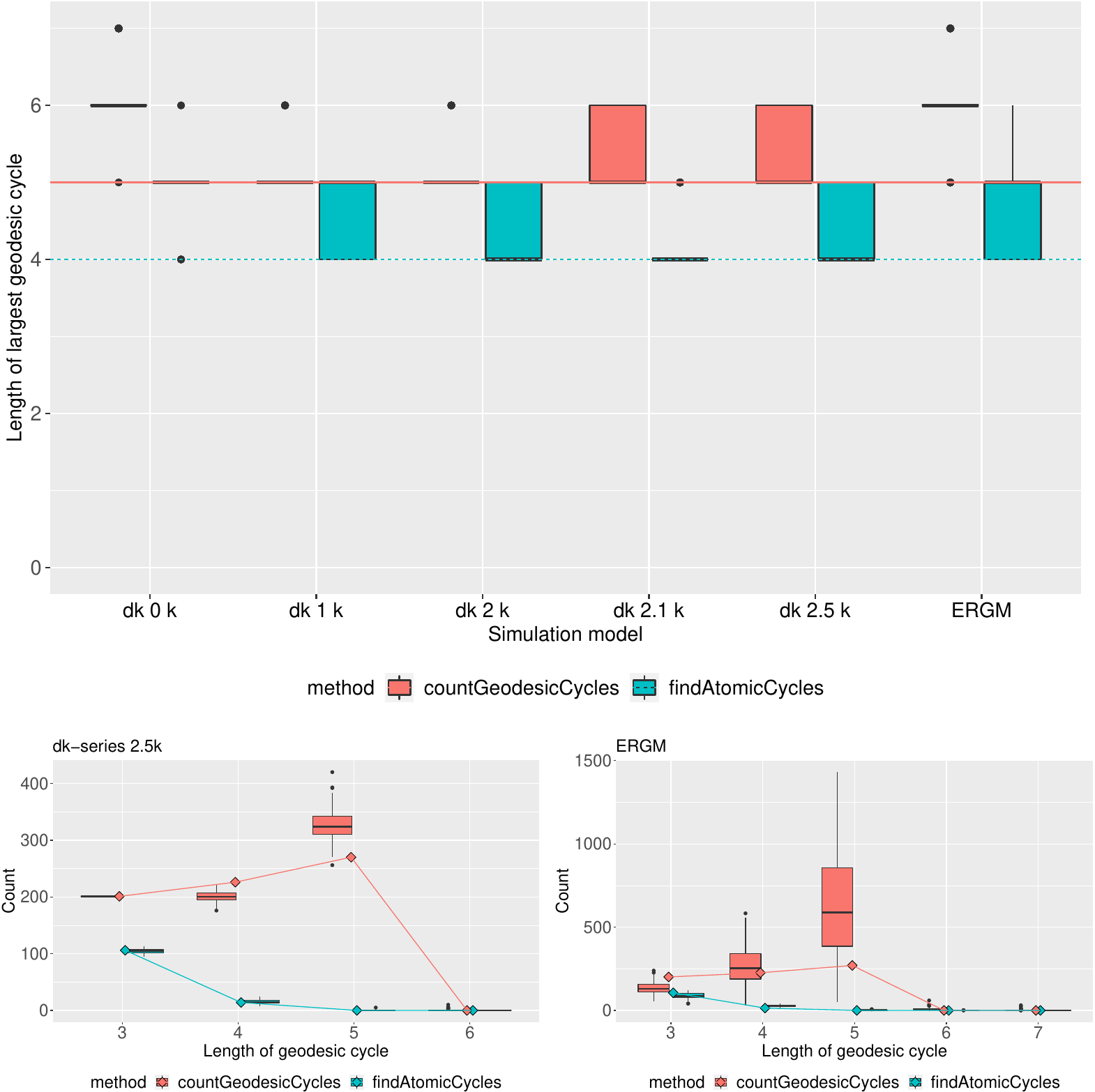}
  \caption{Largest geodesic cycle size (top), and distribution of
    geodesic cycle sizes (bottom) for the Kapferer tailor shop
    network, corresponding to Fig.~13 in \citet{stivala20a}.  The
    bottom two plots show results for the $dk$-series $2.5k$
    distribution (left) and from the ERGM (right).}
  \label{fig:kapferertailorshop_geodesiccycle_lengthdist}
\end{figure}

\begin{figure}
  \centering
  \includegraphics{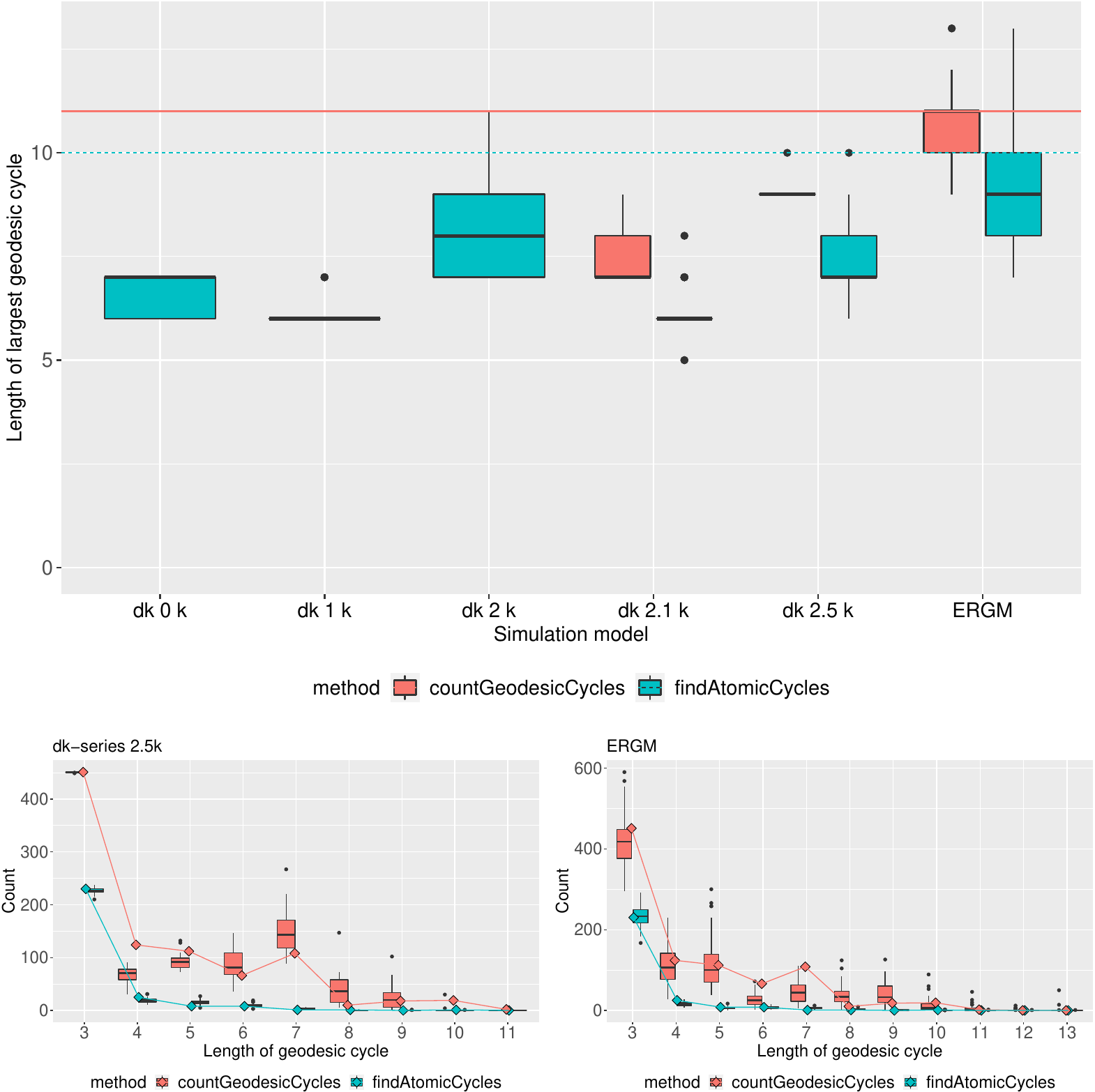}
  \caption{Largest geodesic cycle size (top), and distribution of
    geodesic cycle sizes (bottom) for the high school friendship
    network, corresponding to Fig.~14 in \citet{stivala20a}.  The
    bottom two plots show results for the $dk$-series $2.5k$
    distribution (left) and for the ERGM (right).  These plots
    represent only the 93, 17, and 55 networks (from the 100 simulated
    for each) for dk 2.1 k, dk 2.5 k, and ERGM, respectively, for
    which the cycles could be counted within the 48 hour time limit.
  }
  \label{fig:highschoolfriendship_geodesiccycle_lengthdist}
\end{figure}

\newpage
\setcounter{table}{0}
\setcounter{figure}{0}
\section{ERGM models}
\label{sec:models}

\begin{table}[ht]
  \caption{ERGM model of the Dumbledore's Army network.}
  \label{tab:dumbledoresarmy_ergm}
  \begin{center}
\begin{tabular}{lrrrl}
\hline
Effect & Estimate & std. error & p-value & \\
\hline
edges & $-3.0729$ & $0.4863$ & $<0.0001$ &  *** \\
gwdeg.fixed.0.5 & $1.0245$ & $1.0997$ & $0.351507$ &   \\
gwesp.fixed.0.5 & $0.7965$ & $0.2297$ & $0.000526$ &  *** \\
\hline
\end{tabular}
  \end{center}
  \parbox{\textwidth}{\footnotesize Note: *** $p < 0.001$, ** $p < 0.01$, * $p < 0.05$.}  
\end{table}

\begin{table}[ht]
  \caption{ERGM model of the Harry Potter peer support network.}
  \label{tab:harrypotter_ergm}
  \begin{center}
\begin{tabular}{lrrrl}
\hline
Effect & Estimate & std. error & p-value & \\
\hline
edges & $-3.0046$ & $0.4370$ & $<0.0001$ &  *** \\
nodefactor.gender.2 & $-0.0823$ & $0.1920$ & $0.66839$ &   \\
nodefactor.house.2 & $-0.6507$ & $0.2420$ & $0.00716$ &  ** \\
nodefactor.house.3 & $-1.0031$ & $0.2391$ & $<0.0001$ &  *** \\
nodefactor.house.4 & $-1.7512$ & $0.2927$ & $<0.0001$ &  *** \\
nodefactor.schoolyear.1986 & $1.0385$ & $1.0957$ & $0.34326$ &   \\
nodefactor.schoolyear.1987 & $-0.3320$ & $0.3851$ & $0.38868$ &   \\
nodefactor.schoolyear.1988 & $-0.1574$ & $0.7422$ & $0.83203$ &   \\
nodefactor.schoolyear.1989 & $-0.1282$ & $0.2341$ & $0.58389$ &   \\
nodefactor.schoolyear.1990 & $-0.5506$ & $0.3505$ & $0.11619$ &   \\
nodefactor.schoolyear.1992 & $0.2222$ & $0.3519$ & $0.52778$ &   \\
nodefactor.schoolyear.1993 & $-1.6409$ & $0.6248$ & $0.00863$ &  ** \\
nodefactor.schoolyear.1994 & $-2.4096$ & $0.5896$ & $<0.0001$ &  *** \\
nodematch.gender & $-0.0533$ & $0.2448$ & $0.82777$ &   \\
nodematch.house & $2.1908$ & $0.2767$ & $<0.0001$ &  *** \\
nodematch.schoolyear & $2.0168$ & $0.3120$ & $<0.0001$ &  *** \\
\hline
\end{tabular}

  \end{center}
  \parbox{\textwidth}{\footnotesize
    Note: *** $p < 0.001$, ** $p < 0.01$, * $p < 0.05$.}
\end{table}

\begin{figure}
  \centering
  \includegraphics[angle=270,width=\textwidth]{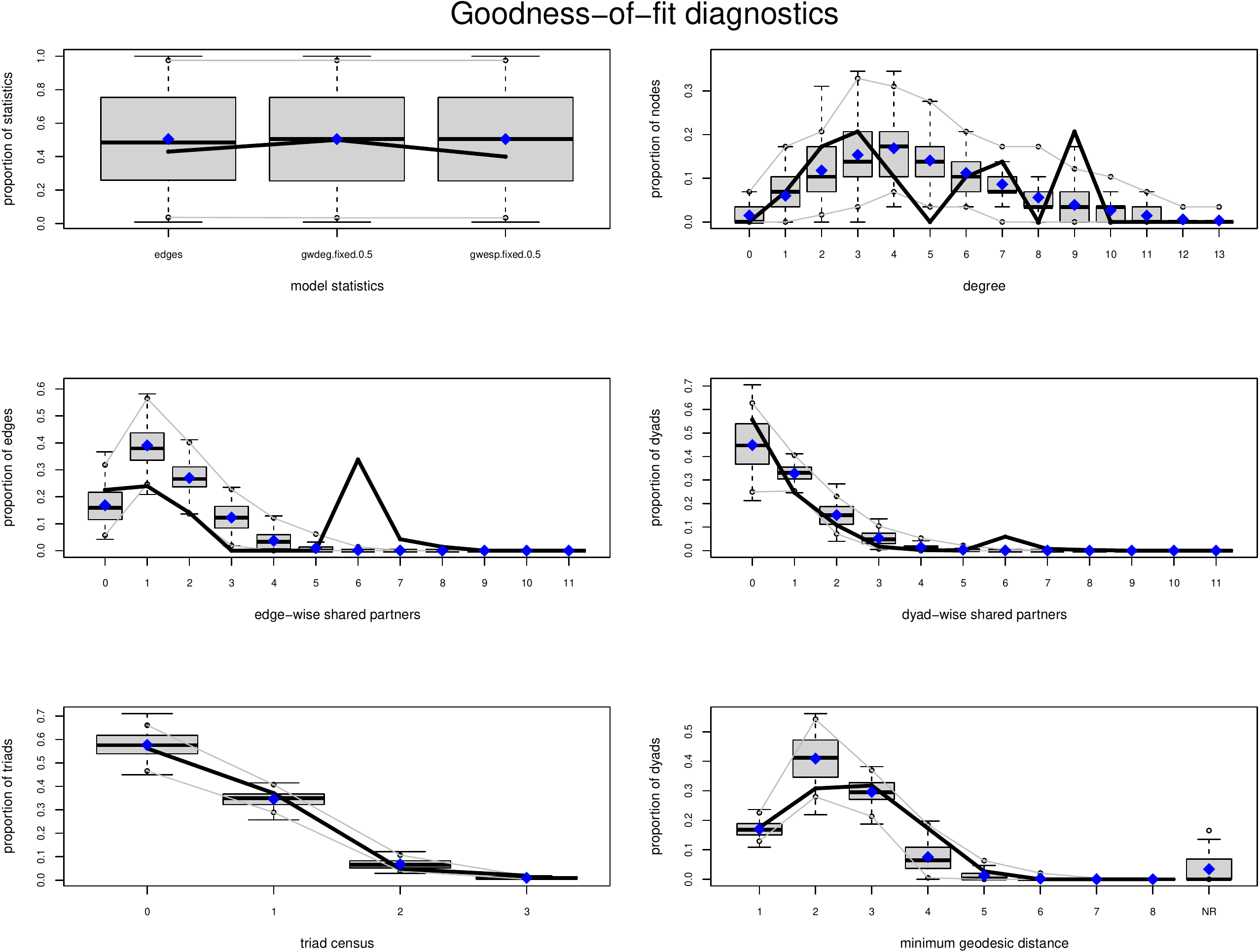}
  \caption{Statnet goodness-of-fit plot for the Dumbledore's Army ERGM
    in Table~\ref{tab:dumbledoresarmy_ergm}.}
  \label{fig:dumbledoresarmy_statnet_model1_gof}
\end{figure}

\begin{figure}
  \centering
  \includegraphics[angle=270,width=\textwidth]{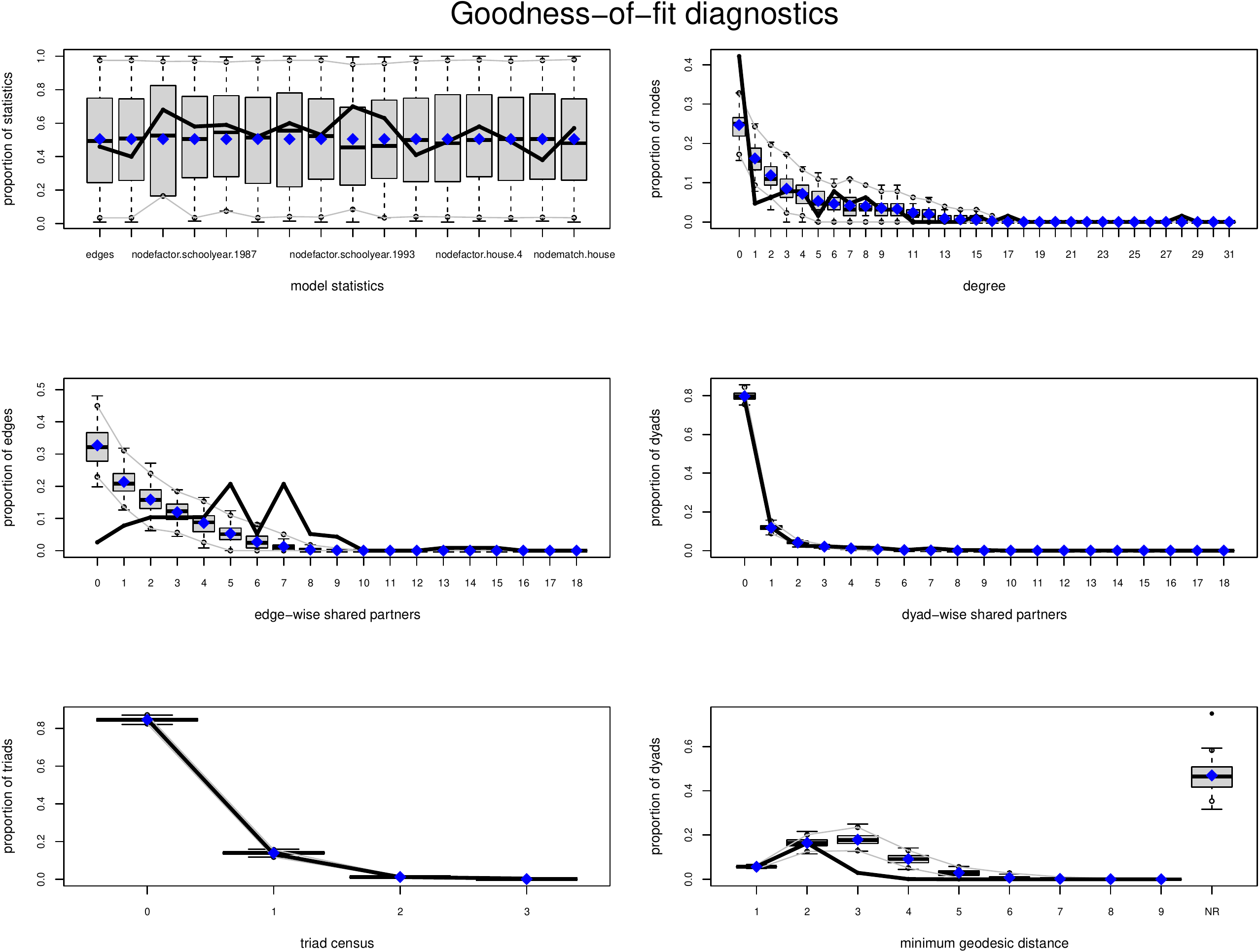}
  \caption{Statnet goodness-of-fit plot for the Harry Potter peer
    support network ERGM in Table~\ref{tab:harrypotter_ergm}.}
  \label{fig:harrypotter_statnet_model1_gof}
\end{figure}

\newpage


\begin{thebibliography}{64}
\bibitem[Bossaert and Meidert(2013)]{bossaert13}
G.~Bossaert and N.~Meidert.
\newblock ``{We} are only as strong as we are united, as weak as we are
  divided'' a dynamic analysis of the peer support networks in the {Harry
  Potter} books.
\newblock \emph{Open Journal of Applied Sciences}, 3\penalty0 (2):\penalty0
  174--185, 2013.
\newblock \doi{10.4236/ojapps.2013.32024}.
\end{thebibliography}

\end{document}